%% file: main.tex
\documentclass[%
 reprint,
superscriptaddress,
 amsmath,amssymb,
 aps,
justification=justified
]{revtex4-2}

\usepackage{graphicx}
\usepackage{bm}
\usepackage{hyperref}
\usepackage[mathlines]{lineno}
\usepackage{url}
\usepackage{color}
\usepackage{graphicx}
\usepackage{mathtools}
\usepackage{float}
\usepackage{gensymb} 
\usepackage{indentfirst}
\usepackage{mathrsfs}
\usepackage{cuted}
\usepackage{multirow}
\usepackage[utf8]{inputenc}

\newcommand{\trm}{\textrm}

\renewcommand{\figurename}{Fig.}

\begin{document}

\preprint{APS/123-QED}

\title{Constraints on the Diffuse Flux of Ultra-High Energy Neutrinos \\ from Four Years of Askaryan Radio Array Data in Two Stations}

\input{ara_revtex_institutes.tex}
\input{ara_revtex_authors.tex}

\date{\today}

\begin{abstract}
The Askaryan Radio Array (ARA) is an ultra-high energy (UHE, $>10^{17}$\,eV) neutrino detector designed to observe neutrinos by searching for the radio waves emitted by the relativistic products of neutrino-nucleon interactions in Antarctic ice. In this paper, we present constraints on the diffuse flux of ultra-high energy neutrinos between $10^{16}-10^{21}$\,eV resulting from a search for neutrinos in two complementary analyses, both analyzing four years of data (2013-2016) from the two deep stations (A2, A3) operating at that time. We place a 90\% CL upper limit on the diffuse all flavor neutrino flux at $10^{18}$ eV of $EF(E)=5.6\times10^{-16}$~$\textrm{cm}^{-2}$$\textrm{s}^{-1}$$\textrm{sr}^{-1}$. This analysis includes four times the exposure of the previous ARA result, and represents approximately 1/5 the exposure expected from operating ARA until the end of 2022.

\end{abstract}

\keywords{Radio-Cherenkov; high-energy neutrinos}
\maketitle

\section{Introduction}

Ultra-high energy neutrinos (UHE, $>10^{17}$\,eV) are a unique window on the distant, high energy universe.
In addition to gravitational waves, they are the only Standard Model messengers capable of traveling cosmic distances undeflected and unattenuated.
Cosmic rays have their trajectories bent by magnetic fields, and for sources more distant than ${\sim}50$\,Mpc, above ${\sim}10^{19.5}$\,eV cosmic rays are expected to be degraded in energy through interactions with the Cosmic Microwave Background (CMB) via the Greisen-Zatsepin-Kuz'min (GZK) effect~\cite{Greisen1966, Zatsepin:1966jv}. Cosmic ray nuclei are additionally degraded in-flight to earth through their natural beta and inverse-beta decay processes, as well as photo-disintegration e.g.\ the Giant Dipole Resonance~\cite{Berman:1975tt}.
High-energy gamma rays ($\gtrsim100$\,TeV) are similarly expected to pair-annihilate off the CMB and Extragalactic Background Light (EBL)~\cite{Gould1967}.

Predictions for the sources of very high energy neutrinos fall broadly into two classes. First, \textit{astrophysical neutrinos} are expected from the site of cosmic ray acceleration, for example gamma ray bursts and active galactic nuclei \cite{Waxman:1999ai,Murase:2015ndr}.
The IceCube experiment has confirmed the existence, and measured the spectrum, of TeV-PeV astrophysical neutrinos \cite{Aartsen:2015knd}, and has identified a first potential source in the blazar TXS 0506+056~\cite{IceCube:2018dnn,IceCube:2018cha}. Second, \textit{cosmogenic neutrinos} are expected from the destruction of cosmic rays through the aforementioned processes~\cite{Beresinsky:1969qj}. A more complete discussion of how the flux of cosmogenic neutrinos depends on the primary cosmic-ray composition, and the effects of various interaction and decay processes, can be found in the literature~\cite{Hooper:2004jc,Allard:2006mv,Kotera:2010yn,Kotera:2011cp,vanVliet:2017obm}.

At energies above above $10^{16}$\,eV, low predicted fluxes \cite{Ahlers:2012rz, Thomas:2017dft} combined with small expected cross sections \cite{Connolly:2011vc,CooperSarkar:2011pa} lead to $\mathcal{O}(10^{-2})$\,neutrino interactions per cubic-kilometer of ice per year per energy decade.
As such, the active volumes of the instruments required to detect this UHE flux must necessarily approach the scale of $100$\,km$^3$ water equivalent. Several experiments are operating or under-construction to reach this high energy flux, including IceCube~\cite{Aartsen:2018vtx}, Pierre Auger~\cite{Aab:2019auo}, NuMoon~\cite{NuMoon}, ANITA~\cite{Gorham:2019guw}, ARIANNA~\cite{Anker:2019rzo}, GRAND~\cite{Alvarez-Muniz:2018bhp}, and ARA~\cite{Allison:2011wk}, which is the focus on this work.

\begin{figure*}[htp]
\centering
\includegraphics[width=0.75\textwidth]{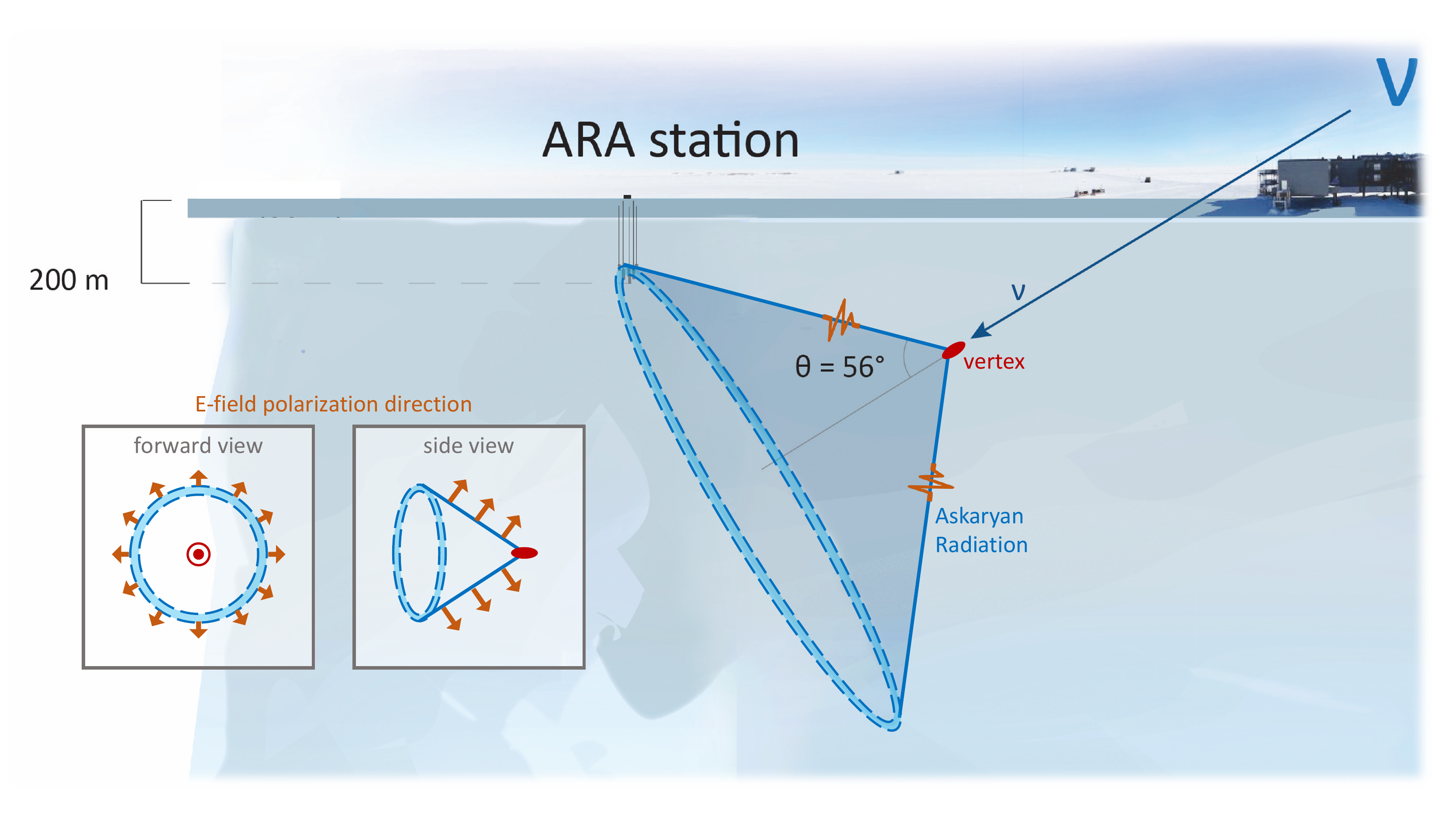}
\caption{A diagram showing how a high energy neutrino interaction might be observed in an ARA station. The insets show how the Askaryan emission and its polarization would be observed if seen along, and perpenduclar to, the shower axis. A more detailed view of an ARA station can be found in Fig.~\ref{fig:ara5_layout}.}
\label{fig:interaction_diagram}
\end{figure*}

The Askaryan Radio Array (ARA) is a UHE neutrino detector deployed at the South Pole seeking to observe these ultra-high energy neutrinos. ARA searches for neutrinos by looking for the broadband (few hundred MHz to few GHz) radio impulse, or ``Askaryan emission"~\cite{Askaryan:1962hbi, Askaryan:1965}, that accompanies neutrino-nucleon interactions.
This effect, caused by a $\sim20\%$ negative charge asymmetry in electromagnetic showers in media, and acting as a coherently radiating current distribution, has been observed in the laboratory at accelerator facilities~\cite{Gorham:2006fy}.
The radiation has a Cherenkov-like beam pattern, with a cone thickness of a few degrees. The leading edge of the electric field pulse points toward the shower axis.
Experiments looking for Askaryan radiation are deployed in dielectric media such as ice, salt, and sand, which are expected to be sufficiently transparent to radio waves as to make the radio signal observable. In the case of ARA, the long (generally greater than 500\,m \cite{barwick2005south}) attenuation length of radio waves in South Pole ice allows naturally occurring detector volumes to be instrumented sparsely and economically.
A diagram of how a neutrino interaction might be observed in an ARA detector is given in Fig.~\ref{fig:interaction_diagram}.

In this paper, we report constraints on the diffuse flux of UHE neutrinos over the energy interval $10^{16}-10^{21}$\,eV. This result is based on two complementary searches for neutrinos in four years of data from ARA stations A2 and A3 recorded between February of 2013 and December of 2016.
This paper is organized as follows.
In Sec.~\ref{sec:instrument_description}, we describe the ARA instrument.
In Sec.~\ref{sec:data_analysis_methods}, we describe the data analysis methods used in two parallel analyses, and in Sec.~\ref{sec:results} we discuss our findings. In Sec.~\ref{sec:systematics}, we discuss systematic uncertainties. Finally, Sec.~\ref{sec:discuss} we discuss the result and its implications, as well as prospects for the future.
We also include an appendix, App.~\ref{app:limit_calc}, where we discuss the calculation of our limit and detail the livetime of the instrument.

\section{Instrument Description}
\label{sec:instrument_description}

\begin{figure*}[htp]
\centering
\includegraphics[width=0.49\textwidth]{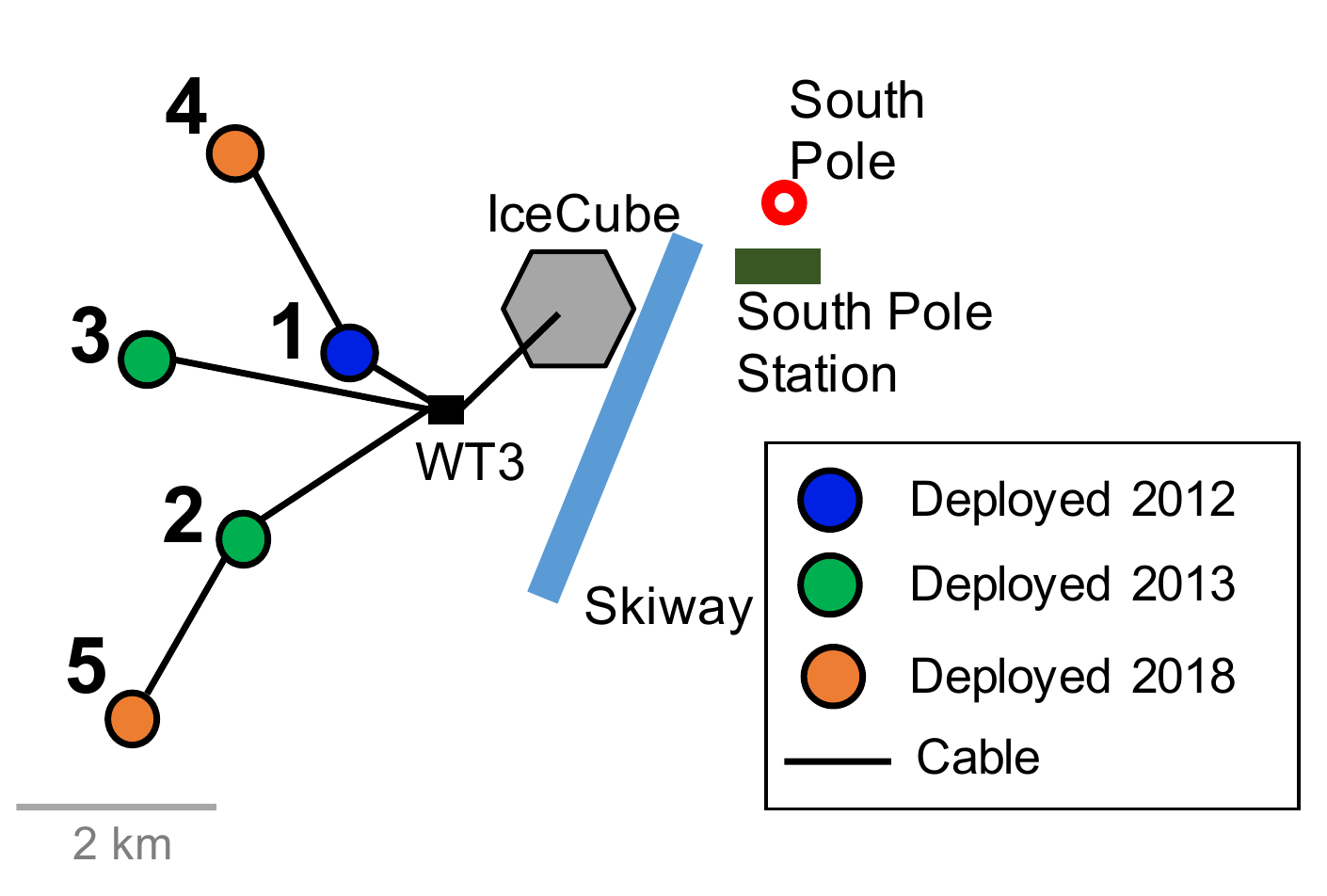}
\includegraphics[width=0.49\textwidth]{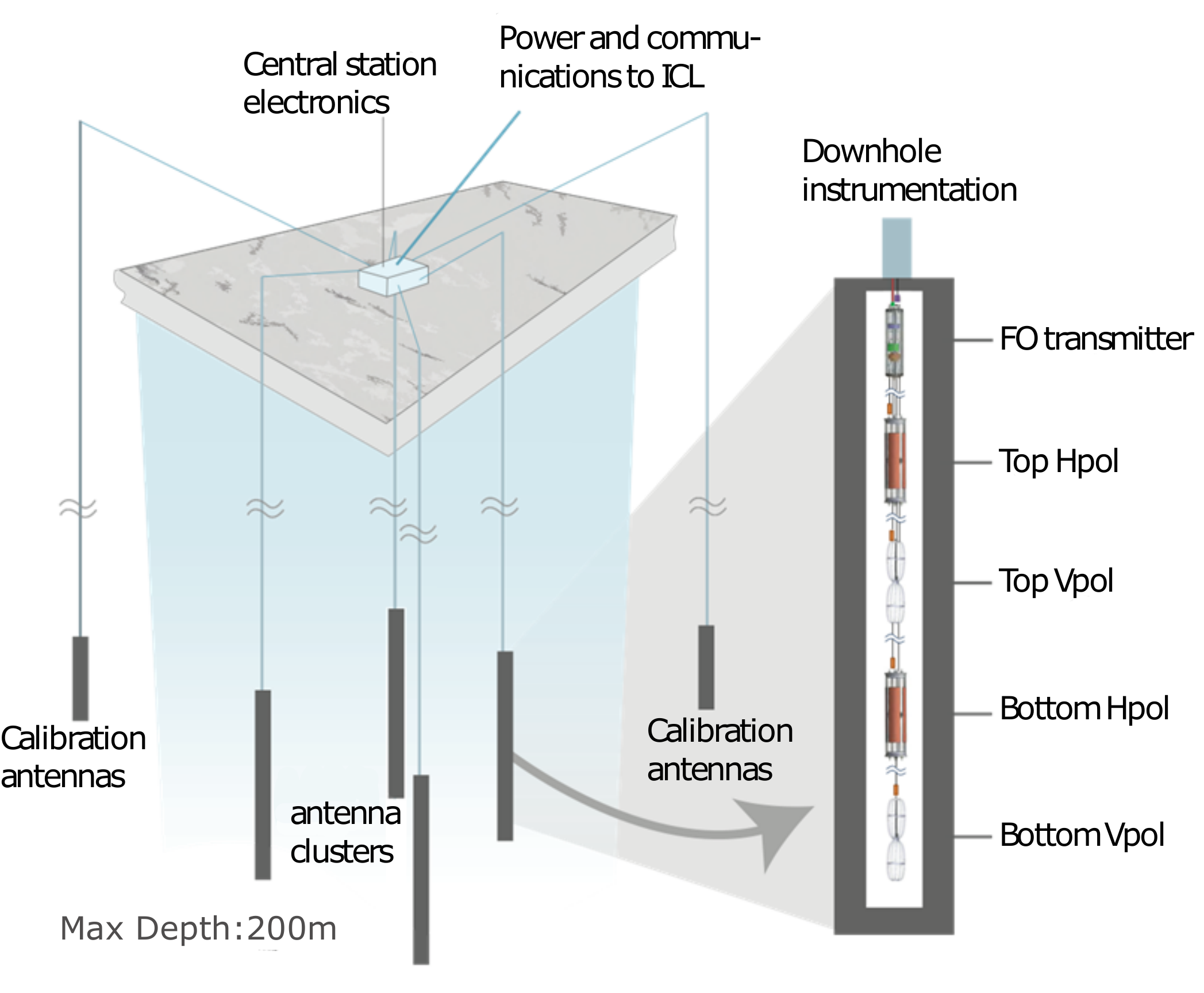}
\caption{(Left) A top-down view of the ARA5 instrument as deployed at the South Pole, with stations color-coded by the year they were deployed. The green stations, A2 and A3, are the focus of the analysis described in this paper. (Right) A schematic of the electronics and instrumentation in an ARA station; ``FO" is a fiber-optic transmitter.}
\label{fig:ara5_layout}
\end{figure*}

The Askaryan Radio Array is a UHE radio neutrino detector consisting of five stations located a few kilometers grid-west of the geographic South Pole in Antarctica, as drawn in Fig.~\ref{fig:ara5_layout}~\cite{Allison:2011wk}.
A single station consists of 16 antennas, eight for detecting horizontally-polarized (HPol) radiation and eight for detecting vertically-polarized (VPol) radiation, along with signal conditioning and Data Acquisition (DAQ) electronics.
The antennas are deployed at the bottom of holes at up to 200\,m depth on four ``measurement strings,'' forming an rectangular solid 20\,m tall and 15\,m deep and wide.
At each corner of the rectangle an HPol quad-slotted cylinder antenna sits a few meters above a VPol wire-frame bicone antenna.
Each antenna is approximately sensitive to radiation in the 150-850\,MHz band~\cite{Allison:2011wk}.
Two ``calibration strings" are deployed about 40\,m radially away from the center of the station.
Each calibration string contains a VPol and an HPol antenna, and is capable of emitting broadband RF pulses, which provide an \textit{in-situ} calibration of station geometry and timing, as well as a measurement of livetime.

Construction of ARA began in 2011, when a prototype station (Testbed) was deployed \cite{Allison:2011wk, Allison:2014kha} at 30~m depth to evaluate the RF environment and electronics performance.
The first design station (A1) was deployed in 2012, but only up to 100\,m depth due to limited drill performance.
In 2013, two deep stations (A2, A3) that are the focus of this work were deployed at up to 200\,m depth~\cite{Allison:2015eky}. Two more 200~m depth stations (A4, A5) were deployed in 2018.

\subsection{The ARA Electronics}
\label{sec:ara_elec}

A schematic drawing of the ARA instrumentation and electronics is shown in the right of Fig.~\ref{fig:ara5_layout}. After an incoming signal excites an antenna, it enters an antenna-mounted front-end signal-conditioning module; there, the signal undergoes a strong ($>50$\,dB) notch filter at 450~MHz to remove South Pole Station communications, is band-passed between 150-850\,MHz, and boosted by approximately 40~dB through two stages in low Low-Noise Amplifiers (LNAs).
The signal is then transmitted
to the surface via RF-over-Fiber (RFoF) to reduce attenuation over the 200\,m journey to the top of the borehole.
At the surface, the optical signal is converted back to an electronic signal, amplified again by $40$\,dB, before finally being bandpass filtered once more to remove any out-of-band noise contributed by the amplifiers.
The signal is then split into two paths, one for triggering and one for digitization.

The trigger path is routed through a tunnel-diode which serves as a passive, few-nanosecond power integrator.
When the rising edge of the tunnel diode output exceeds roughly five times the ambient thermal noise level, the lowest-level single channel trigger fires.
If three same-polarization antennas register a single channel trigger within 170~ns (the light propagation time in the ice across the station's diagonal) all 16 antennas in the station are read out.
This scheme is optimized to trigger on Askaryan pulses, which should generate significant power in very short time windows
and traverse the array at the speed of radio propagation in ice (${\sim}0.16$m/ns).

\begin{figure}[ht]
\centering
\includegraphics[width=\columnwidth]{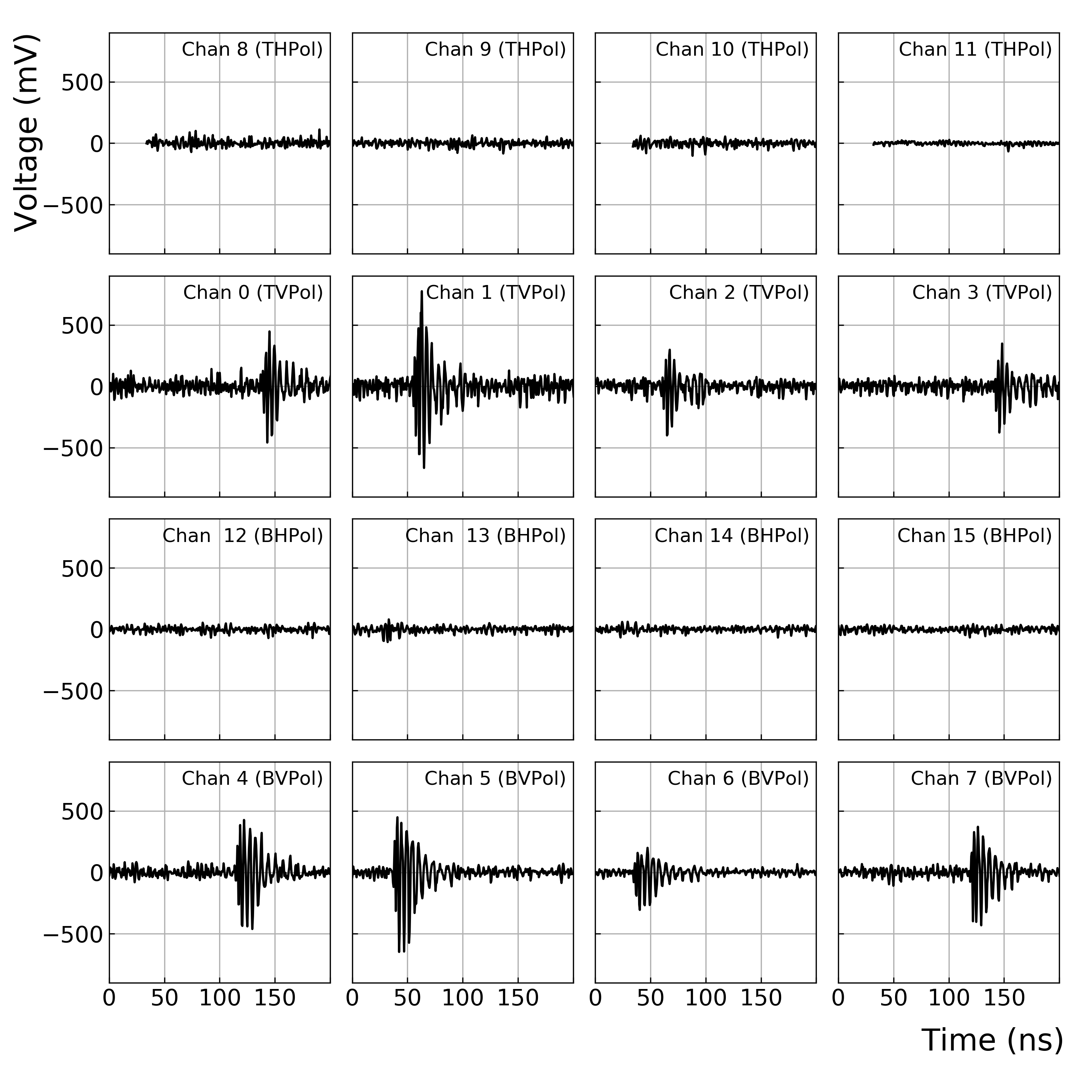}
\caption{An event display showing the sixteen waveforms recorded in A2 for a VPol calibration pulser.}
\label{fig:event_display}
\end{figure}

The signal is recorded through the digitization path.
The signal is stored in the circular buffer of an IceRay Sampler 2 (IRS2) chip, which is a high-speed 3.2\,Gs/s digitizer ASIC~\cite{VarnerIRS2}.
To minimize power consumption, the buffers are implemented in analog as Switched Capacitor Arrays (SCA)~\cite{Varner:2007zz,Roberts:2018xyf}.
After a global trigger is issued, sampling is halted and analog-to-digital conversion commences.
Each readout records 400-600\,ns of waveform, roughly centered on the trigger.
The bundle of 16 waveforms and the associated housekeeping data (UTC timestamp, etc.) defines an \textit{event}. An example VPol calibration pulser event is shown in Fig.~\ref{fig:event_display}, where ``TVPol" notes a vertically-polarized antenna deployed at the top of a string, ``BHPol" notes a horizontally-polarized antenna deployed at the bottom of a string, and so forth.

Triggering is performed by four Triggering DAughter boards (TDAs), while digitization is handled by four Digitizing DAughter boards (DDAs), with four RF channels per board.
The logic and readout to storage for the eight daughter boards is managed by the ARA Triggering and Readout Interface (ATRI).
The ATRI communicates via USB with a Linux Single Board Computer (SBC) for run control and data archiving. A more detailed discussion of the ARA electronics can be found in previous work~\cite{Allison:2011wk,Allison:2015eky}.

The precise triggering threshold for a given antenna is adjusted to maintain a single channel trigger rate for that antenna. The targeted single channel rates are chosen so that the global trigger rate, after taking into account combinatorics and trigger coincidence windows, is maintained at 5\,Hz.
The dominant source of these ``RF triggers" is fluctuations in the blackbody thermal noise background of the ice, but also includes any potential neutrino signals, as well as anthropogenic (human-made) signals such as aircraft, motor vehicles, etc.
In addition, each station collects a sample of background ``software'' internally-generated triggers as well as the calibration pulses, both at 1\,Hz, for a total 7\,Hz global trigger rate. Every triggered event invokes approximately 1\,ms of deadtime in the electronics readout system,  which has negligible $<1$\% impact on the livetime.

\subsection{Detector Livetime}
\label{sec:livetime}
\begin{figure*}[ht]
\centering
\includegraphics[width=\linewidth]{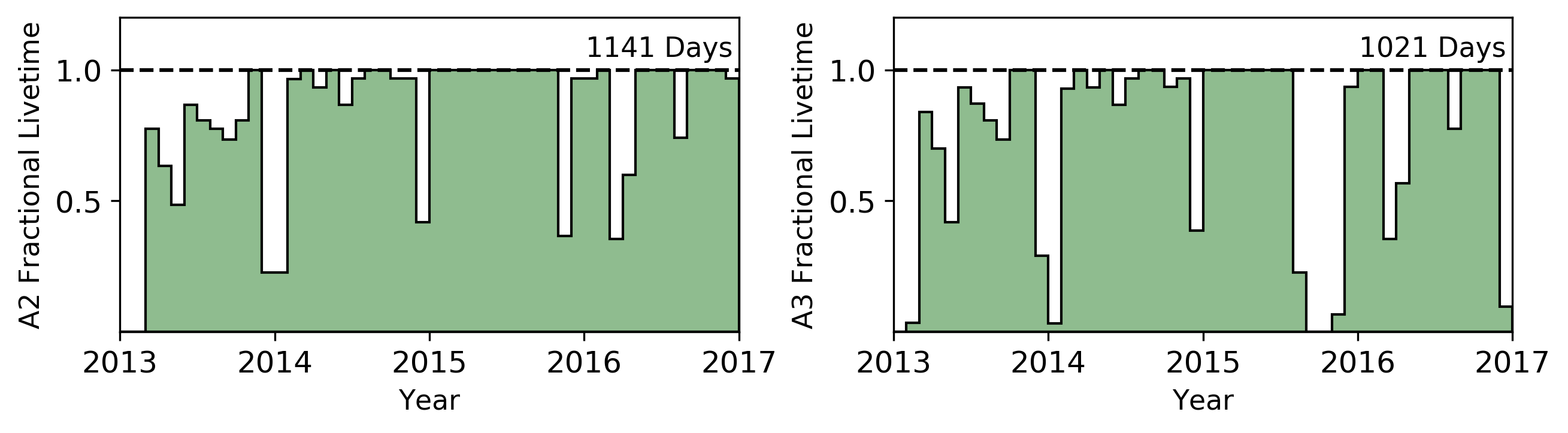}
\caption{Operational fractional livetimes for A2 (left) and A3 (right) from deployment in February 2013 through the end of the analysis period in 2016; each bin is one month wide. From the 4\,years of deployment, 1141 days from A2, and 1021 days from A3, are good for analysis. This is mostly due to intermittent downtime; quality cuts remove less than 2\% of livetime.
}
\label{fig:A23_uptime}
\end{figure*}

This analysis comprises data recorded by ARA Stations 2 and 3 (A2 and A3) between initial deployment in February 2013 and the end of December 2016. Over the course of these four years, each station accumulated roughly 1100\,days of livetime, as shown in Fig~\ref{fig:A23_uptime}, recording over 1.2 billion events total between the two stations. The two detectors were operated in several different ``configurations", representing different combinations of operating parameters such as trigger window size, etc. We summarize the five data taking configurations for each station in 
Tab.~\ref{tab:configs} of App.~\ref{app:livetime}. For all configurations in A2, the bottom HPol channel of string 4 was non-operational, and it is excluded from participating in the trigger for configurations 3-5. Additionally, for configurations 3-5 of A3, the fourth string of the detector
participates in forming triggers normally, but due to technical problems in the digitization chain it does not produce useful signal for analysis.

\subsection{Simulation}
\label{sec:simulation_description}
We generate simulated data sets with the Monte Carlo package \texttt{AraSim}, which has been previously described extensively in Allison \textit{et al.} \cite{Allison:2014kha, Allison:2015eky}.
This code models the generation of neutrino events from a diffuse flux and their interactions with Earth and Antarctica.
After simulating interactions in-ice, \texttt{AraSim} renders a time-domain parameterization of the Askaryan radiation and propagates that radiation through the ice, taking into account signal attenuation and ray bending based on a depth-dependent index of refraction model.
When the radiation arrives at a simulated station, it is convolved with a frequency-dependent model of the detector, including the antennas, signal chain, and the trigger logic. The model of the instrument includes the dispersive effect of the signal chain that induces a frequency-dependent group delay. 
If the event satisfies a simulated trigger, it is stored in the same format as real data so that our analysis codes can be executed on either data or simulated
events interchangeably.

The models of the A2 and A3 stations are data-driven, and include calibrations derived from the 2012-2013 dataset as described in \cite{Allison:2015eky}.
In particular, the antenna locations, the noise temperature of the ice, and the gain of every channel are all implemented in the model based on \textit{in situ} measurements. The simulation also models the configuration-specific variations in the electronics behavior (readout length, trigger window size, trigger delay values, etc.) as detailed in App.~\ref{app:livetime}.

\begin{figure}[ht]
\centering
\includegraphics[width=\columnwidth]{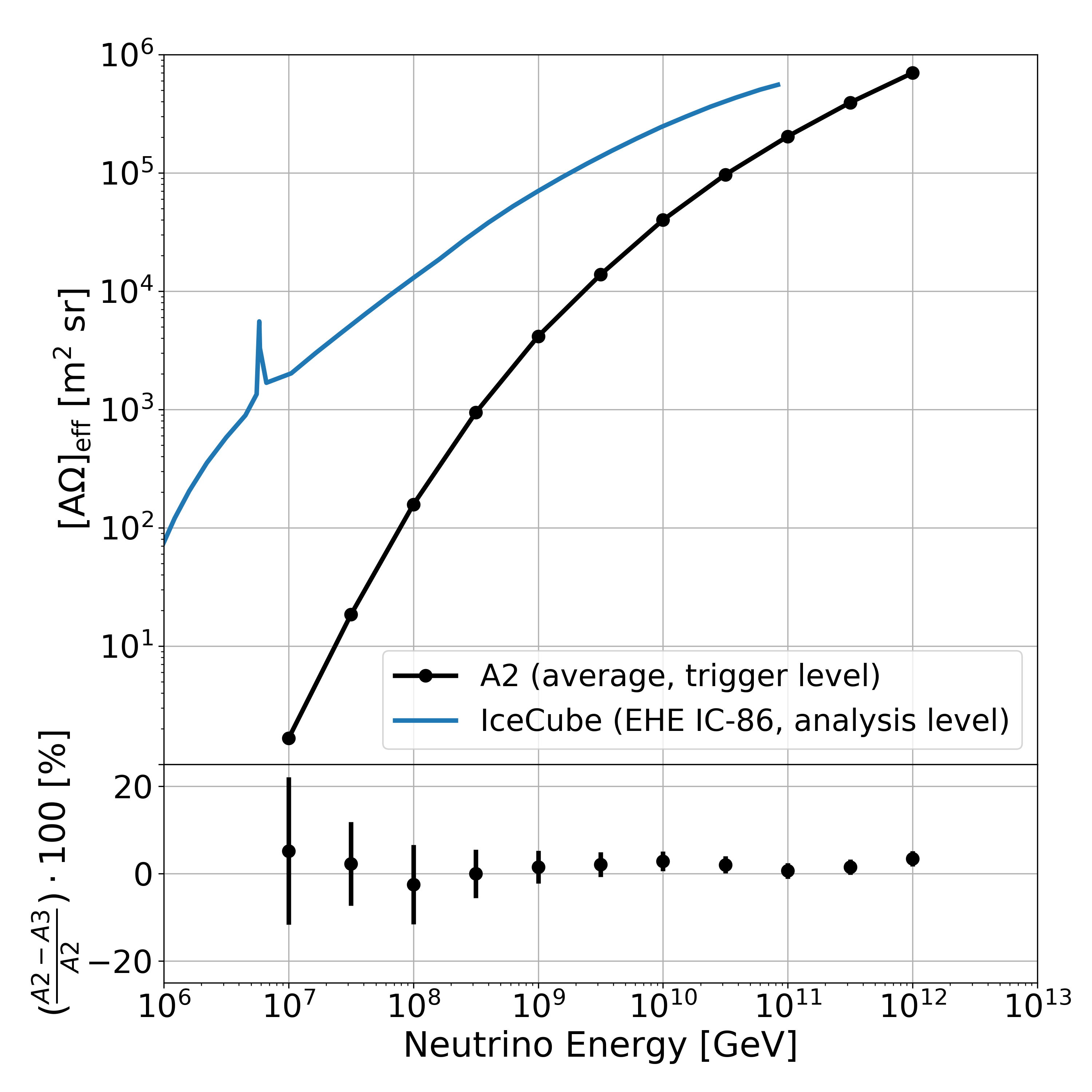}
\caption{(Top) The simulated trigger-level effective area-steradian ($[A\Omega]_{\trm{eff}}$) for A2, averaged across configurations. For comparison, we also show the analysis-level sensitivity of IceCube~\cite{Aartsen:2013dsm}. (Bottom) The percent difference between the A2 and A3 effective areas.}
\label{fig:aeff}
\end{figure}

In Fig.~\ref{fig:aeff}, we show the aperture ($[A\Omega]_{\rm{eff}}$) of A2, averaged over configurations. The effective area is derived via Monte Carlo techniques with \texttt{AraSim} as described in App.~\ref{app:limit_calc}. For comparison, we also plot the effective area of the IceCube experiment~\cite{Aartsen:2013dsm}. As can be seen in the bottom panel, we find that A2 and A3 have comparable effective areas to within a few percent. We additionally find that triggering and readout parameters specific to each livetime configuration, as discussed in Sec.~\ref{sec:livetime}, do not result in differences in the trigger level effective area in excess of a few percent. The two detectors, A2 and A3, are simulated independently; previous studies have shown that only a small fraction of events trigger both A2 and A3 simultaneously, amounting to about 5\% of events at 1 EeV ~\cite{Allison:2015eky}.

\section{Data Analysis}
\label{sec:data_analysis_methods}
Our data analysis searches for a diffuse flux of neutrinos between $10^{16}{-}10^{21}$\,eV. 
The analysis is designed to remove background events, principally thermal and anthropogenic noise, while preserving sensitivity to neutrinos.
The analysis proceeds in a ``blind'' fashion, where the ARA data is divided into two subsets.
A ``burn'' sample of 10\% of the data, which is assumed to be representative of the full data sample, is set aside and used to tune cuts and understand backgrounds.
The remaining 90\% of the data is kept blinded 
until cuts are finalized.
Before unblinding, it was decided that in the absence of a detection, the analysis with the best \textit{expected} limit would be used to set the limit.

\subsection{Summary of Blind Analyses}

Two parallel, complementary analyses were performed on the four-year data samples, which we refer to as Analysis A and Analysis B. In this section, we outline the strategies followed by both, with Sec.~\ref{sec:analysis_highlights} describing features specific to the two separate analyses.

Both analyses follow similar strategies. First, a set of basic data quality and livetime cuts are applied to remove detector `glitches', calibration events, and periods of livetime known to be contaminated with anthropogenic activities. Second, fast event-level filters designed to reduce the quantity of data by an order of magnitude or more are applied. Third, interferometric-based reconstruction is performed to identify the arrival direction of a recorded signal, and geometric cuts invoked to reject events that originate from above the ice surface or in the direction of known calibration pulsers. Finally, a bivariate cut is applied on the signal strength and a reconstruction quality parameter. Analysis A considers events only in the vertical polarization, while Analysis B includes events in both the horizontal and vertical polarizations. Both derive data-driven models of the background, and set their final cuts such that $\sim 0.01$ background events are expected to pass the analysis in the 1100 days of livetime, corresponding to the level at which we find the best expected limit.

Both analyses use the 10\% ``burn" sample to tune cuts and understand backgrounds. The number of neutrinos expected in the burn sample based on allowed models is $<0.02$, so the chance of excluding a neutrino candidate in the burn sample is negligible. Moreover, cuts were motivated by an optimization procedure and by examining outlying events that did not have neutrino-like properties, not by the objective to eliminate all events in the burn sample. The distribution of the background was smooth and followed a statistical distribution, and we did not readjust cuts to remove specific neutrino-like events.

\subsection{Data Quality Cuts}

Before analysis begins, we remove periods of livetime which are known to contain human and calibration activity. This includes, for example, maintenance operations on the detector during the Austral summer, and the operation of surface pulsers or pulsers deployed on two of the final IceCube strings (strings 1 and 22), known as the ``IceCube Deep Radio Pulsers." These livetime cuts remove less than 2\% of the total livetime recorded by the instrument.

Next, both analyses deploy a nearly common set of data quality cuts designed to remove instrumental glitches and remaining calibration events from the dataset. Glitches are typically present either as waveforms that are shorter than those generated during normal readout, or waveforms with unphysical discontinuities (likely due to digital errors in the readout electronics), and comprise less than 0.001\% of events.
Additionally, we remove the internally-generated ``software" triggers described above in Sec.~\ref{sec:ara_elec}, as well as ``tagged" calibration pulser events. We are able to ``tag" calibration pulsers under normal operating conditions as they are nominally triggered by the pulse-per-second (PPS) TTL inside the DAQ, so these events are readily identified by their timestamps. This has negligible affect on the detector livetime and neutrino sensitivity. The quality cuts that are not common between analyses focus on slightly different methods for detecting unphysical discontinuities in the waveforms, as well as the identification of out-of-band power content.

\subsection{Event Filter}
Because of the large size of the ARA dataset (${>1.5\,\times\,10^{8}}$\,events/station/year), and the expectation that most triggers are upward fluctuations of the thermal noise environment, each analysis applies a computationally simple cut that rejects $>$90\% of triggered events.  Analysis A utilizes an event filter based on a multiplicity condition, which requires that more than three VPol channels each have an signal strength above a threshold. Analysis B utilizes a wavefront-RMS filter, which requires that the pattern of arrival times across the array be consistent with that of a plane-wave. Both algorithms have been described elsewhere \cite{Lu:2017amt}. Analysis A tunes its filter such that 99\% of triggered events do not pass the filter, while Analysis B tunes its filter such that approximately 90\% of triggered events do not pass. In Analysis A, the signal strength threshold is tuned. In Analysis B, the signal strength threshold and tolerance parameter for deviation from plane wave-like timing is tuned.
In Analysis~A, the multiplicity trigger is approximately 70\% efficient for $10^{18}$\,eV neutrinos, where for Analysis~B the wavefront-RMS filter efficiency is approximately 90\%.

\subsection{Reconstruction and Geometric Cuts}
\label{sec:reco}
For events passing the event filter, we perform an interferometric-based reconstruction to determine the direction of the source of measured incoming radio waves. This interferometric reconstruction technique has been used in other ARA analyses~\cite{Allison:2014kha, Allison:2015eky, Allison:2015lnj, Allison:2018whu} and in the ANITA experiment~\cite{Romero-Wolf:2014pua}.  The interferometric technique relies on the relationship between the location of an emitting source in space and the time delays expected for two measurement antennas with known separation.

For a given pair of antenna waveforms, the cross-correlation $C_{\trm{i,j}}$ between the voltage waveform on the $i$-th antenna ($V_{\trm{i}}$) and the voltage waveform on the $j$-th antenna ($V_{\trm{j}}$) as a function of time lag $\tau$ can be expressed in Eq.~\ref{equ:crosscorr}:
\begin{equation}
C_{i,j}(\tau)=\frac{\sum\limits_{t}V_i(t)V_j(t+\tau)}{RMS_i \times RMS_j}
\label{equ:crosscorr}
\end{equation}
where the $RMS$ are the root-mean-square voltages of the waveforms in the absence of signal. The lag $\tau$ defines the the time delay of one antenna waveform relative to the other and depends on the position of the source emitter relative to the array center, characterized by an elevation angle ($\theta$), an azimuthal angle ($\phi$), and a distance to the source ($R$). The array center is defined as the centroid of all sixteen measurement antennas in the station. 

The pairwise time lags $\tau$ for a given point on the sky $\theta, \phi$ are computed by calculating the path a light ray would take from a hypothesized source located at a distance $R$ to an antenna. The calculation accounts for the changing index of refraction of the Antarctic firn, which causes rays to follow curved, rather than rectilinear trajectories. With $n(z)$ the depth-dependent index-of-refraction, and $z$ the (negative) depth from the ice surface, the ray-tracing method models the changing index of refraction as:
\begin{equation}
n(z)=1.78 - 1.35 e^{0.0132z}.
\end{equation}
This index of refraction model was determined by fitting data collected by the RICE experiment in Antarctica~\cite{kravchenko_besson_meyers_2004}. We consider the index to be unity above the surface.

The total cross-correlation strength for a given point on the sky is given by summing over all like-polarization pairs of antennas as in Eq.~\ref{eq:skycorr2}:
\begin{equation}
C_{\rm{sky}}(\theta,\phi; R)=\frac{\sum_{i=1}^{n_{ant}-1}\sum_{j=i+1}^{n_{ant}}C_{i,j}[\tau(\theta,\phi; R)]}{n_{\rm{ant}}}
\label{eq:skycorr2}
\end{equation}
To smooth uncertainties in the ice model and other systematics (such as differences in the phase responses of the various contributing antennas),  we calculate the Hilbert envelope of the cross-correlation function before summing over pairs, as is done in previous analyses. The Hilbert envelope of the cross-correlation $H(C_{\trm{i,j}})$ is calculated according to Eq.~\ref{eq:hilbert}:
\begin{equation}
H(C_{i,j}) = \sqrt{C_{i,j}^2 + h^2({C_{i,j}})}
\label{eq:hilbert}
\end{equation} 
where $h(C_{\trm{i,j}})$ denotes the Hilbert transform.

The cross-correlation function for an individual pair of antennas, $C_{\trm{i,j}}$, is expected to be maximal when the lag is equal to the true difference in the arrival times of a signal at the two different antennas. The sky map is therefore expected to have a peak at the putative source direction.

For determining source direction, Analysis~A tests radii from 40-5000\,m to locate a hypothesis radius which maximizes $C_{\trm{sky}}$, while Analysis~B reconstructs only at 41\,m and 300\,m, corresponding to the radius of the calibration pulser and a radius taken as a plane-wave proxy. That one analysis performs a radius scan is a setup inherited from a separate investigation regarding our ability to determine the radius of curvature for signals, which we found to be limited for sources beyond a few hundred meters, given the instrumental timing resolution. After finding the best reconstruction direction (the direction which maximizes $C_{\trm{sky}}$), both analyses impose two geometric cuts. The first is an angular cut in the direction of the calibration pulser system. The second is a cut on the reconstructed elevation ($\theta$) of the hypothetical source relative to the station center, and is used to reject events coming from above the surface.

The cuts on the angular region around the calibration pulser systems is necessary to reject untagged calibration pulser events; approximately 1 in $10^{4}$ calibration pulser signals are emitted outside of the time window expected; the cause of this ``misfiring" is not well understood. Additionally, one configuration in A3 (configuration 2) did not have the calibration pulser system correctly synchronized to the PPS clock, and so a purely geometric rejection criterion is needed. To determine this geometric cut region, the angular distribution of tagged calibration pulsers is fit (either with a Gaussian or a Kernel Density Estimator), and a cut region determined such that fewer than $10^{-3}$ calibration pulser events are expected to reconstruct outside of that angular region for the entire livetime period. The angular cut region is an approximately $10^{\circ}\times10^{\circ}$ box around the true calibration pulser location. The value of $10^{-3}$ is approximately an order of magnitude less than the number of background events expected to pass all analysis cuts. Less than 3\% of neutrinos are cut by this calibration pulser geometric cut requirement.

The geometric cut at the surface is used primarily to reject anthropogenic noise, as well as potential downgoing physics signals such as cosmic rays. We make the cut on events from above the surface because we expect neutrino events to predominantly yield up-coming signals. The cut on events from above the surface proceeds similarly to the calibration pulser geometric cut. We fit the distribution of events in $\sin(\theta)$ near the transition between the air-ice boundary, and place an angular cut such that fewer than $10^{-3}$ events from the above-ice distribution are expected to reconstruct within the ice. In Analysis~A, events are only reconstructed in the vertical polarization, while in Analysis~B, an event may be classified as having an above-the-surface origin in either polarization, and if so it is rejected from consideration in the searches in either polarization. The cut on the reconstruction angle $\theta$ varies from $11$-$35^{\circ}$, and approximately 10-30\% of neutrinos are cut by the surface cut at $10^{18}$~eV, depending on the analysis, station, and configuration. For example, in Analysis A, the angular cut is $\sim30^{\circ}$ for A2, but is $\sim10^{\circ}$ for A3. The reduction in efficiency is partially because radio waves can follow curved trajectories as they traverse the varying index-of-refraction, and can appear as downgoing signals when they in fact arise from sources within the ice.

\subsection{Bivariate Cut and Background Estimate}

\begin{figure*}[ht]
\centering
\includegraphics[width=0.49\linewidth]{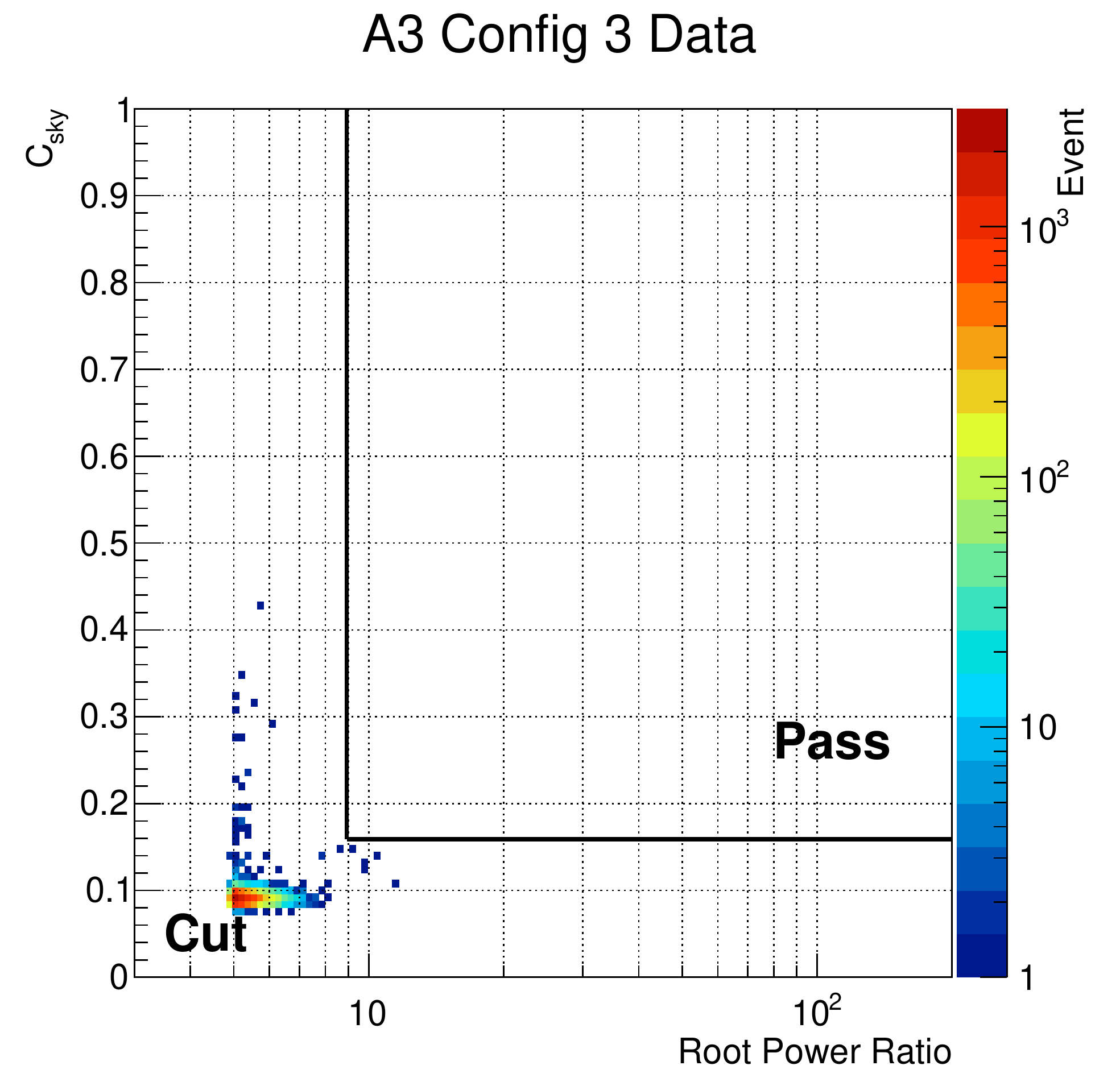}
\includegraphics[width=0.49\linewidth]{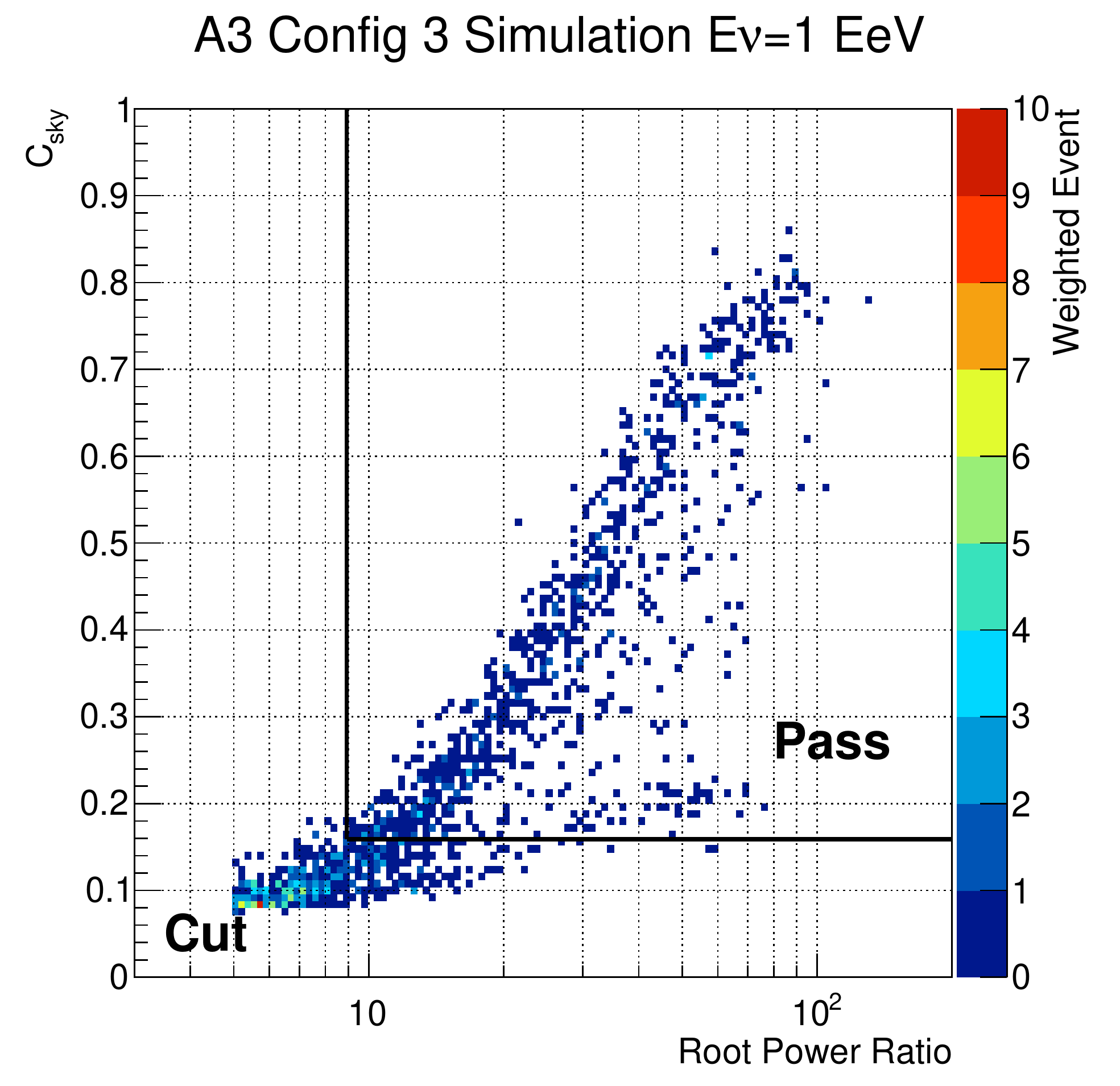}
\caption{An example of the bivariate cut plane, for which the final 2-D box cut is made for A3 configuration 3. (Left) The plane as observed in 10\% ``burn-sample" data, showing events clustering at low-correlation and low-root-power-ratio. (Right) The plane populated with simulated neutrinos at $10^{18}$\,eV, showing events distributed throughout. Events at low-correlation and low-root-power-ratio are cut; events at higher values define the signal region, and pass the analysis.}
\label{fig:LDcut}
\end{figure*}

Both analyses implement their final separation of noise from potential neutrino signals as a bivariate cut in the peak cross-correlation ($C_{\trm{sky}}$) vs.\ signal strength ($\Gamma$) plane. For an event to ``pass," Analysis~A imposes a box cut requiring that an event's $C_{\trm{sky}}$ and $\Gamma$ both exceed a station and configuration specific threshold: $C_{\trm{sky}}>C_{\trm{min}}$ and $\Gamma>\Gamma_{\trm{min}}$. In Analysis~B, an event is required to pass a linear combination of the two, such that $\Gamma > m \, C_{\trm{sky}} + b$, where $m$ and $b$ are station- and configuration- specific analysis parameters. An example of the box cut for A3 configuration 3, in Analysis~A, is provided in Fig.~\ref{fig:LDcut}.

For the purpose of Fig.~\ref{fig:LDcut}, we show $\Gamma$ in the way it was computed to perform cuts in Analysis~A. This definition of signal strength we call the root-power-ratio (RPR), and is defined as $RPR=E_{\trm{j,max}}/\sigma_{E_{\trm{j,noise}}}$, where $E_{\trm{j,max}}$ is the maximum of the square-root of a rolling 25\,ns integrated power average of the waveform, specifically:
\begin{equation}
\label{eq:RPR}
E_j = \sqrt{\frac{1}{n} \sum_{i=j}^{j+n} V_i^2}
\end{equation}
where $n$ is the number of samples in the 25\,ns window and $\sigma_{E_{\trm{j,noise}}}$ is the RMS value of $E_{\trm{j}}$ in the half of the waveform that does not contain the maximum. This RPR variable has been used in a previous ARA analysis~\cite{Allison:2015eky}, and was chosen to more-closely emulate the power-integrated envelope that is used in the ARA trigger.

Both analyses use a data-driven model of the backgrounds in order to set final cuts and estimate the expected number of background events passing all cuts. As in previous analyses, the model is constructed by fitting the distribution of events as a function of the cut parameters ($C_{\trm{sky}}$, RPR, etc.), and setting the cut such that fewer than $\sim0.01$ background events are expected to pass all cuts, which is the level at which we find the best expected limit based on statistical uncertainties only.

Before examining the neutrino signal region, defined as events passing all cuts in the analysis, both analyses first reversed the requirement that events reconstruct inside the ice. That is, we examined events which failed the geometric cut by reconstructing to the surface. This is done in order to identify bursts of activity from the surface, and we exclude runs which have $\gtrsim11$ events reconstructing to the surface. At this stage, we do not exclude single, isolated events, ``singlets," as neutrinos are expected to arrive isolated in time and space. In both analyses, this ``surface-noisy" cut eliminated approximately an additional week of livetime.

\subsection{Analysis-Specific Comments}
\label{sec:analysis_highlights}
\subsubsection{Analysis A}
Analysis A uses solely signal from VPol antennas for the search. This is motivated by the fact that the majority ($\sim 70$\%) of simulated signal events contain VPol triggers. This is partly because VPol antennas are more sensitive than HPols antennas, especially at low frequencies. To define the surface geometric cut, Analysis A reconstructs the incident angle of each event with signal arrival times calculated assuming a bulk-ice model with a constant index of refraction 1.76 and a putative source distance of 5\,km to emulate a distant source at the ice surface. This approach proved to be the most successful in reconstructing a radio emitter system installed on the rooftop of the IceCube Lab, which served as a proxy for distant surface signals. The cut is then placed on the elevation angle of the result of this reconstruction as described in Sec.~\ref{sec:reco}. 

One category of background present in ARA data is continuous-wave (CW) emission. CW emission is anthropogenic in origin and presents as a strong spectral peak in the power spectral density of an event. The most common type of CW encountered in ARA is generated by the $\sim$403\,MHz radiosonde attached to NOAA weather balloons that are launched once or twice daily from the South Pole; we additionally see 125\,MHz emission from an as-yet unidentified source.

To eliminate the contamination of CW emission, Analysis~A places an out-of-band cut, where an event is considered CW-contaminated if more than three channels in either polarization demonstrate peak spectral density below 170~MHz. This frequency threshold is motivated by the edge of the pass-band filter. We discard the event entirely if such CW contamination is found. This cut represents negligible signal efficiency loss below $10^{19}$\,eV, and a $\sim 10$\% loss at $10^{21}$\,eV from off-cone signals. To reject CW contamination in higher frequencies, we observe that such events, while producing high $C_{\trm{sky}}$ values due to their CW nature, do not produce high RPR values on the $C_{\trm{sky}}$-RPR plane. Therefore, Analysis~A rejects these events with the 2-D box cut. 

\subsubsection{Analysis B}
Analysis B features two major differences from Analysis~A. First, Analysis~B performs the neutrino search in both polarizations, VPol and HPol. Second, Analysis B filters power in events with CW contamination. 
CW contamination is identified with two methods: first by looking for spectral peaks over run-specific baselines as in the prototype station analysis~\cite{Allison:2014kha} and second by looking for stability between phasors at a given frequency as is done in the LOFAR experiment~\cite{Schellart:2013bba}. Once CW has been identified at a specific frequency, this contamination is removed using a filtering technique developed and used by the ANITA collaboration which operates in a similar frequency domain~\cite{DaileyThesis,Allison:2018cxu}. The filter notches spectral peaks in the amplitude domain, while reconstructing the phasors representing the signal and thernal noise contributions only, with CW contamination removed.
Once an event has been filtered of its contaminating CW emission, it proceeds in the analysis as above.

Development and use of techniques to mitigate CW contamination is important because the $\sim$403\,MHz emission at South Pole can contaminate up to 10\% of ARA's daily livetime.  As the detectors continue to accrue livetime, and sensitivity to weak signals improves, the ability to filter events of contaminating CW emission will be important for leveraging the full livetime of the array.

\begin{figure*}[ht]
\centering
\includegraphics[width=0.49\linewidth]{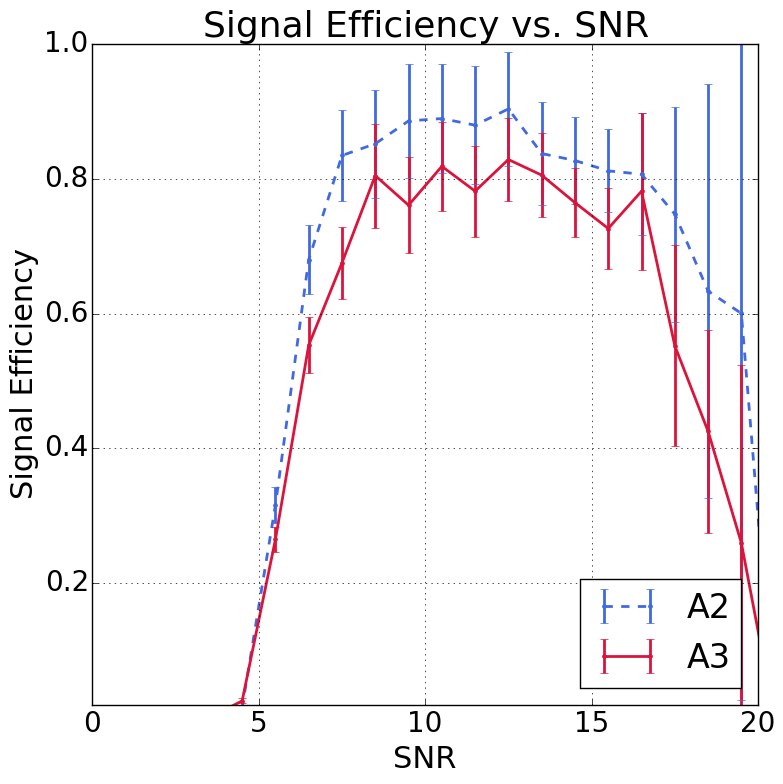}
\includegraphics[width=0.49\linewidth]{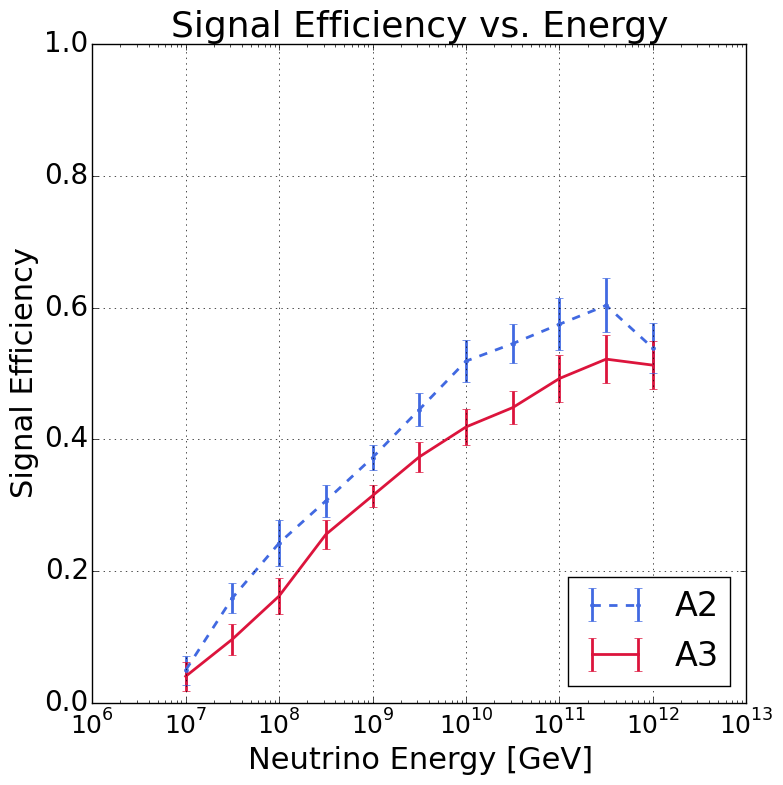}
\caption{Monte-Carlo estimated analysis efficiency as a function of signal-to-noise ratio (left) and neutrino energy (right) for Analysis A. For context, the trigger efficiency of an ARA station has been measured to reach 50\% at an SNR of 3.7~\cite{Allison:2018ynt}. In the left figure, we assume an unbroken power-law spectrum with a spectral index of -2.13 to weight the energies contributing to the efficiency. The efficiency decrease around SNR=14 is due to waveform saturation effects as simulated in \texttt{AraSim}.}
\label{fig:efficiency}
\end{figure*}

\section{Results}
\label{sec:results}

After rejecting data containing bursts of surface activity, both analyses examined the neutrino signal region in A2 before examining the signal region in A3. Each analyses' individual unblinding results are discussed below in Sec.~\ref{sec:results_analysisA} and Sec.~\ref{sec:results_analysisB}.

Neither analysis observes a statistically significant excess of events, observing zero events against the estimated background. In the absence of detection, in Fig.~\ref{fig:limit} we compute the 90\% confidence level (CL) upper limit on the diffuse flux of neutrinos.

Further details on the upper limit calculation, including inclusion of the systematic uncertainties discussed in Sec.~\ref{sec:systematics}, can be found in App.~\ref{app:limit_calc}. Inclusion of the systematic uncertainties in the limit has an $\mathcal{O}(5\%)$ effect.
We report the limit set by Analysis A, which had slightly superior \textit{expected} sensitivity, by up to 15\%, depending on energy. As a benchmark, the number of events expected to be observed in the analysis ranges from 0.25 for an unbroken extrapolation of the astrophysical neutrino flux as measured by IceCube with a spectral index of -2.13~\cite{Aartsen:2016xlq}, to 0.027 in the case of a cosmogenic neutrino flux where protons make up only 10\% of cosmic ray primaries~\cite{Ahlers:2012rz}.

In Fig.~\ref{fig:efficiency}, we present the analysis efficiency of Analysis~A for both A2 and A3; we plot the average signal efficiency, taking into account the variations due to different run configurations and their respective livetimes. The signal efficiency is calculated by simulating neutrinos in AraSim, and taking the ratio of the number of neutrinos passing the analysis cuts to the number of neutrinos that trigger the detector.
We show the efficiency for Analysis~A, as it is the analysis used to set our limit, though the efficiencies for Analysis~B (which was developed in a parallel and independent fashion) are comparable. On the left, we show the efficiency as a function of SNR, where SNR is computed as the third highest $V_{peak}/RMS$, where $V_{peak}$ is the highest absolute voltage peak in a waveform, and the RMS is the root-mean-square of the voltage values in that waveform. We present the figure with this definition of SNR as it more closely aligns with that commonly used for comparison purposes in the literature.
The analysis becomes efficient near an SNR of 6, and does not fully saturate to a value between 75-90\% until it is above an SNR of 8. The saturated efficiency for A3 is $\sim10$\% lower than for A2 because A3 required a larger angular cut region to reject surface events, as discuss in Sec.~\ref{sec:reco}. On the right, we show the efficiency as a function of energy. At $10^{16}$\,eV, the analysis has a relatively low efficiency of about 5\%. The efficiency rises to $\sim$\,35\% by $10^{18}$\,eV and peaks near $10^{20}$\,eV at between 50-60\%, depending on the specific station. Efficiencies for all stations and configurations are provided in additional Figures in App.~\ref{app:livetime}.

\subsection{Analysis A Results}
\label{sec:results_analysisA}
After post-unblinding examination, Analysis~A observes 0 events on a background expectation of ${(5\pm2)\times10^{-2}}$ background events per station.

At unblinding, Analysis~A observed two events in the candidate neutrino signal region in A2. While both reconstruct inside the ice using an interferometric technique which utilizes all VPol channels of the array, both only have visibly identifiable signals in the bottom row of VPol antennas. When the reconstruction is repeated utilizing only antennas 
where the signal strength exceeds the event filter threshold, both events confidently reconstruct to above the surface. We consider both of these events to be backgrounds of surface origin.

At unblinding, Analysis~A observed four events in the candidate neutrino signal region in A3. Three cluster in time to within a few minutes, and are located in a run which contains a burst of surface noise, but was technically sub-threshold in the ``surface-noisy" cut as described above in Sec.~\ref{sec:results}. The fourth event is reconstructed inside the ice when all VPol channels participate in the interferometry.
Again, if only channels with signal strength above the event filter threshold are considered, the event reconstructs to above the surface. It is therefore determined to be consistent with a background of surface origin.

Since all events observed in Analysis~A can, with currently available tools, be identified to be of surface origin, or cluster in time with bursts of surface activity, we do not consider Analysis~A to have measured any events. The post-unblinding cut necessary to remove the misreconstructed surface events results in a negligible efficiency loss ($\leq 0.25\%$). As Analysis~A provided the better expected limit, we proceed to compute the limit as described in App.~\ref{app:limit_calc} with a observed number of events of zero.

\subsection{Analysis B Results}
\label{sec:results_analysisB}
After post-unblinding examination, Analysis~B observes 0 events on a background expectation of ${(1\pm0.3)\times10^{-2}}$ events per station.

At unblinding, Analysis~B observed 19 events in the candidate neutrino signal region in A2. Of these, seven were ``near-surface" events, and were addressed by more stringent, data-driven surface cuts, as described in Sec.~\ref{sec:reco}. Analysis~B had originally used a geometric argument to determine the value of the surface cut, as opposed to data-driven methods. An additional seven events were of a type not observed in the burn-sample, where an unphysical amount of power was deposited in one or two strings. These were removed with an update to the quality cuts, and the update had negligible impact on the signal efficiency. One event was a calibration pulser that ``mis-fired" during a time when it was not enabled by the software. It was misreconstructed in the 41\,m interferometry, but was correctly reconstructed in the 300\,m radius, and was removed by additionally rejecting events if they reconstructed towards the calibration pulser in either interferometric radii. This additional calibration pulser geometric rejection also had negligible impact on the analysis efficiency. To address the remaining four events, an additional hit-time based reconstruction method, which traces its lineage to the RICE experiment, was added. The method uses a integrated-power envelope (the same as described in in the definition of RPR in Eq.~\ref{eq:RPR}) to identify hit times in the waveforms, and requires at least four waveforms in the event to have an RPR above a threshold of eight. The method then searches for the location of a source emitter ($\theta, \phi, R$) which minimizes the differences between the predicted and observed time delays between channels. With this additional cut, all four of the remaining events in Analysis~B are rejected---one fails to have enough hits to be reconstructed, and the remaining three reconstruct to above the surface.

At unblinding, Analysis~B observed three events in the candidate neutrino signal region in A3. Two cluster in time within a few minutes, and are located in the same run which generated the three passing events in Analysis~A. The third is the same event observed in Analysis~A, which was determined to be downgoing both in Analysis~A through the revised interferometric method described above, and also in Analysis~B independently with the hit-time based reconstruction method. Like in Analysis~A, all three events are determined to be of surface origin, or associated with a burst of surface activity.

Since all events observed in Analysis~B can, with currently available tools, be identified to be of surface origin, or cluster in time with bursts of surface activity, we do not consider Analysis~B to have measured any events. The additional post-unblinding hit-time based reconstruction cut results in no more than an additional 2\% efficiency loss.

\begin{figure}[!htb]
\centering
\includegraphics[width=\columnwidth]{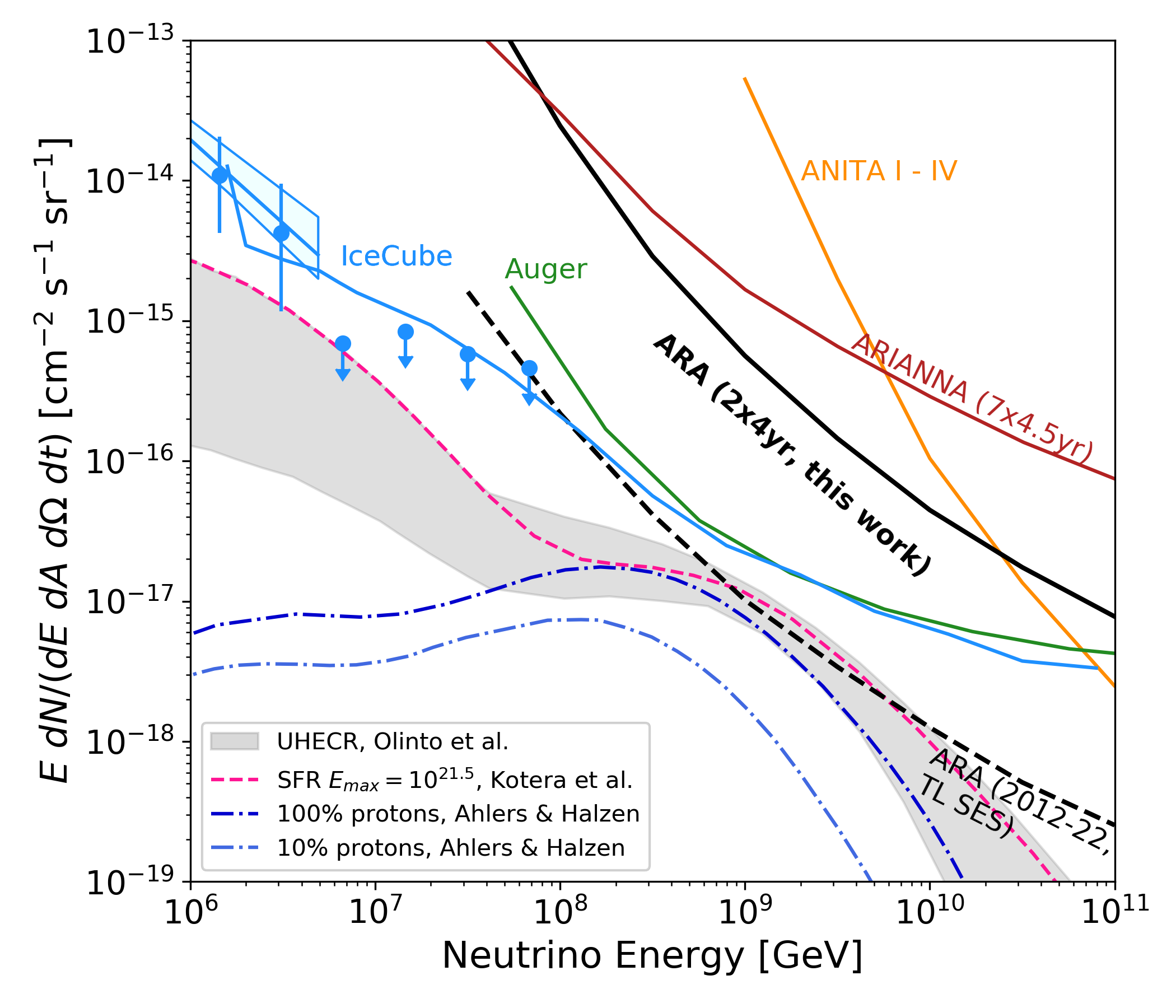}
\caption{The 90\% confidence-level upper limit on the all-flavor diffuse flux of neutrinos set by this analysis (thick black line). The limit accounts for uncertainties in the background estimate and systematic uncertainties on the neutrino sensitivity. We also plot the projected trigger-level single-event sensitivity (TL SES) for the five-station ARA5 array by 2022 as a black-dashed curve. Also shown are the latest limits and flux measurements from IceCube~\cite{Aartsen:2018vtx, Aartsen:2016xlq}, Auger~\cite{Aab:2019auo} (rescaled with decade-wide bins and for all-flavors), ANITA~\cite{Gorham:2019guw} (rescaled with decade-wide bins), and ARIANNA~\cite{Anker:2019rzo}. Shown for comparison are several benchmark cosmogenic neutrino flux models~\cite{Olinto:2011ng,Kotera:2010yn, Ahlers:2012rz}.}
\label{fig:limit}
\end{figure}

\section{Systematic uncertainties}
\label{sec:systematics}

\begin{figure}[!htb]
\centering
\includegraphics[width=\columnwidth]{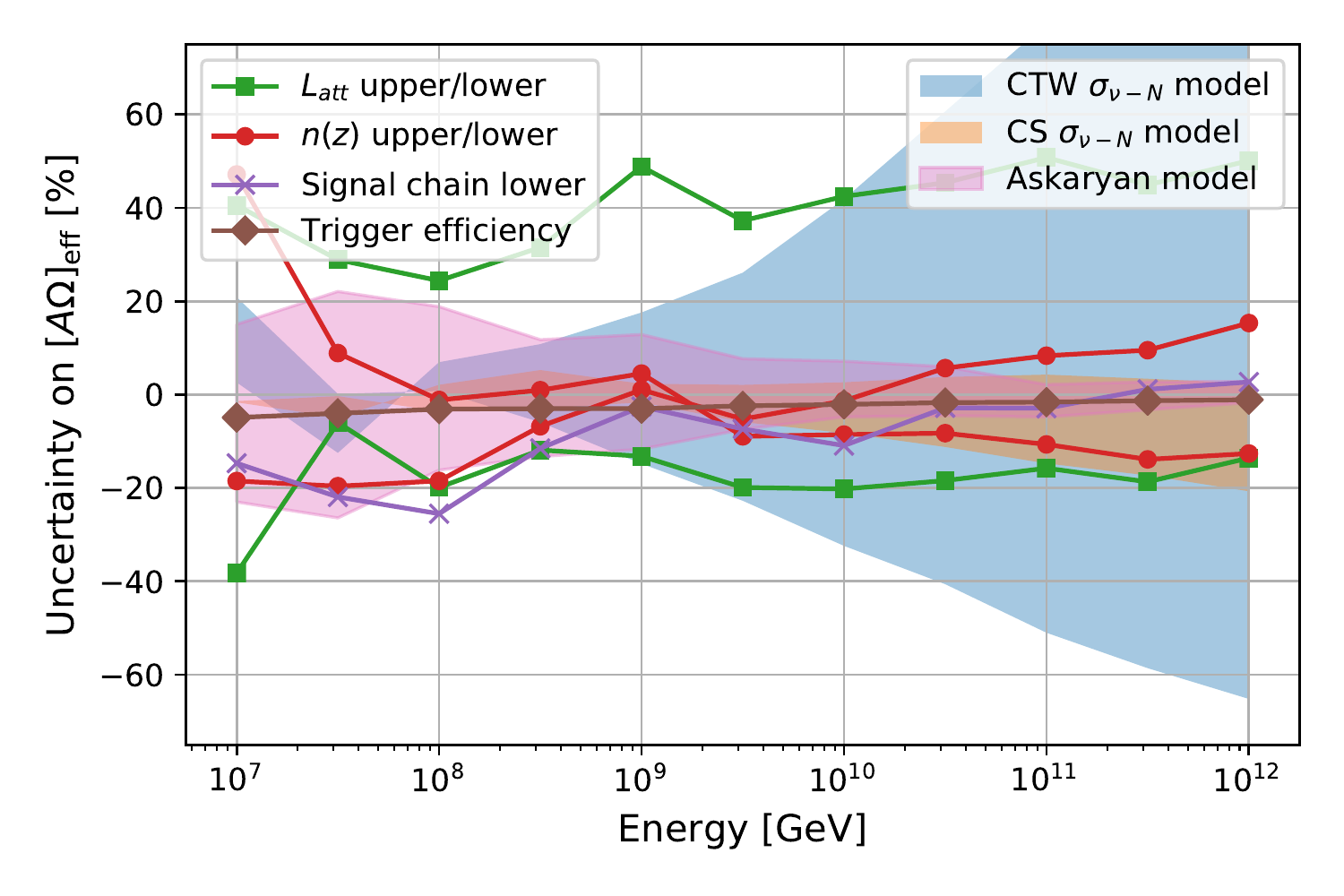}
\caption{Uncertainties between the central values used in the simulation and upper/lower bounds for each model parameters. Theoretical systematics (shaded regions), such as the Askaryan model and the neutrino-nucleon cross section, are not accounted for when calculating the neutrino limit. Uncertainties associated with the detector and medium (dashed and solid lines) are accounted for in the calculation.}
\label{fig:rel_diff}
\end{figure}

In this section, we describe the systematic uncertainties considered in the analysis. The impact of these systematics on  $[A\Omega]_{\trm{eff}}$ are shown in Fig.~\ref{fig:rel_diff}, and a table summarizing the magnitude of their effects at $10^{18}$\,eV is provided in Tab.~\ref{tab:systematic_sizes}. We consider systematic uncertainties broadly in two classes. The first class is associated with theoretical uncertainties surrounding the neutrino-nucleon cross section and Askaryan emission, and are shown in Fig.~\ref{fig:rel_diff} as solid bands, reported at the trigger level. The second class is associated with uncertainties in our understanding of the detection medium and our instrument. The latter are taken into account in setting the final limit as described in App.~\ref{app:limit_calc}, and are shown as dashed/dotted lines in Fig.~\ref{fig:rel_diff} at the analysis level.

For the neutrino-nucleon cross section ($\sigma_{\nu-\trm{N}}$), \texttt{AraSim} uses the model derived by Connolly, Thorne, and Waters (CTW)~\cite{Connolly:2011vc}. The upper and lower bounds for $\sigma_{\nu-N}$ are substituted for the central value in the simulation to estimate the effect of the uncertainty on the simulated $[A\Omega]_{\trm{eff}}$ at the trigger level. In the CTW model, the uncertainties on $\sigma_{\nu-\trm{N}}$ are large and grow as a function of energy, exceeding 100\% above $10^{21}$\,eV. At $10^{18}$\,eV the uncertainties on the trigger-level effective area due to the cross-section are estimated at -15\%/+18\%. In Fig.~\ref{fig:rel_diff}, for comparison we also show the uncertainties if we use an alternative cross-section developed by Cooper-Sarkar~\textit{et. al.}(CS)~\cite{CooperSarkar:2011pa} which has smaller uncertainties at high energies by about a factor of four. 
We additionally studied $d[A\Omega]_{\trm{eff}}/d[\sigma_{\nu-\trm{N}}]$, and find it to be approximately linear; for example, at 1~EeV, a 10\% increase in $\sigma_{\nu-\trm{N}}$ corresponded to a 10\% increase in $[A\Omega]_{\trm{eff}}$.

For the Askaryan emission, \texttt{AraSim} implements a modified version of the model derived by Alvarez-Mu\~niz \textit{et. al.}~\cite{AlvarezMuniz:2011ya}. A full description of modifications is provided elsewhere~\cite{Allison:2014kha}, but the primary differences arise due to the inclusion of of the LPM effect by Alvarez-Mu\~niz but not by \texttt{AraSim}, and in \texttt{AraSim}'s use of functional parameterizations for the shower profile instead of directly simulated shower profiles. The relative difference between waveform amplitudes produced by \texttt{AraSim}, and those derived from a full shower Monte-Carlo are at most ${\sim}12$\%~\cite{HongThesis}. We conservatively estimate the effect of this systematic uncertainty by reducing or increasing the simulated field amplitude by $\pm12\%$ and assessing the change in $[A\Omega]_{\trm{eff}}$ at the trigger level. The relative difference between the default parameterization and the scaled parameterization has a maximum value of about 25\% near $10^{16}$\,eV, and starts falling as energy increases. This is because at high energies the instrument acceptance becomes dominated by geometric effects (ray tracing, etc.) and not signal amplitude.
At $10^{18}$\,eV the estimated uncertainties due to the Askaryan emission model are -11\%/+13\%.

In the second category of uncertainties, we consider those arising from our detector response and from measurements of quantities such as the index of refraction in ice and the attenuation length of radio waves in ice. These systematics are included in our calculation of the final limit. We consider uncertainties associated with (1) the attenuation length ($L_{\trm{att}}$) of South Pole Ice and (2) the depth-dependent index of refraction ($n(z)$) of South Pole ice, (3) the calibration of the ARA signal chain, and (4) the triggering efficiency of the detector.

The model for the attenuation length ($L_{\trm{att}}$) of South Pole ice was derived from data taken with the ARA Testbed prototype~\cite{Allison:2011wk}. Confidence bands providing an upper and lower limit on $L_{\trm{att}}$ are given in the model. To set upper/lower limits on our sensitivity, in \texttt{AraSim}, the upper and lower bounds for $L_{\trm{att}}$ are substituted for the central value.
At $10^{18}$\,eV the uncertainty on the analysis level effective area due to uncertainties in attenuation length are -8\%/+50\%.
 
The model for the depth-dependent index of refraction $n(z)$ was obtained by fitting data obtained by the RICE experiment~\cite{kravchenko_besson_meyers_2004}. The data was fitted with an exponential as a function of (negative) depth $z$ of the form $n_d-(n_d-n_s)e^{z\cdot n_c}$, finding the following parameter values and their respective uncertainties: $n_d=1.788\pm 0.016,$, $n_s=1.359\pm 0.022$ and $n_c =0.0132 \pm 0.0017\, \textrm{m}^{-1}$. We recalculate the sensitivity, setting all parameters to their upper and lower limits simultaneously. The lower (upper) limit generally corresponds to a slower (faster) transition from surface to deep ice, and correspondingly have a smaller (larger) geometric acceptance for neutrinos. Additionally, since we do not change the ice-model assumption used to reconstruct the incoming direction of the RF emission as discussed in Sec.~\ref{sec:reco}, this systematic uncertainty also captures errors which may be present if the true ice model for radio wave propagation does not match that used for reconstruction. At $10^{18}$\,eV the uncertainties on the analysis level effective area due to the index of refraction model are 5\%.

We consider four sources of uncertainties that exist in the signal chain. They are the transmission coefficient  $t$ representing the impedance mismatch between the ice and the antenna, as well as between the antenna and the coaxial cable, the ambient noise power received $N_{\trm{ant}}$, the signal chain noise power $N_{\trm{sc}}$, and the antenna directivity $D$. We follow the treatment used in the previous ARA result~\cite{Allison:2015eky} where we consider the system signal-to-noise ratio representing the ratio of input signal power to total system noise power in a given channel:
\begin{equation}
    SNR_{\rm{sys}} = \dfrac{tDP_{\rm{sig}}}{tN_{\rm{ant}}+N_{\rm{sc}}}
\end{equation}
with $P_{\trm{sig}}$ being the received signal power. The four sources of uncertainty translate to an uncertainty in $SNR_{\trm{sys}}$ by standard error propagation, which is then implemented as an uncertainty in the antenna gain $G$ in code ($\Delta G = \Delta SNR_{\trm{sys}} / P_{\trm{sig}}$). In line with previous ARA work, here we only consider the case where the effective gain of the instrument is reduced, providing a conservative estimate of our sensitivity. This is done because we lack sufficient calibration data at this time to constrain the upper bound on the gain.  The VPol antenna gain has an overall estimated uncertainty of -10\%, while the HPol antenna gain is estimated at -32\%. The modified gain values are substituted in the simulation to assess the impact of this uncertainty, and the uncertainty at $10^{18}$\,eV is found to be -3\%.

For the systematic uncertainty associated with the trigger efficiency of the detector as a function of RPR, $\epsilon(RPR)$, we compare the simulated trigger efficiency $\epsilon_{\trm{sim}}(RPR)$ to the measured trigger efficiency in calibration pulser data $\epsilon_{\trm{dat}}(RPR)$: ${\Delta \epsilon = \epsilon_{\trm{dat}}(RPR) - \epsilon_{\trm{sim}}(RPR)}$. We measure $\epsilon_{\trm{dat}}(RPR)$ by varying a tunable attenuator on the local calibration pulsers described in Sec.~\ref{sec:instrument_description} and counting the number of calibration pulsers recorded. 
Using \texttt{Arasim} we find that the uncertainties on the trigger efficiency decreases the simulated $[A\Omega]_{\trm{eff}}$ from between 2-5\% depending on energy, and at $10^{18}$\,eV the size of the effect is 
-3\%.

We observed in previous calibration exercises that the stations trigger inefficiently on calibration pulsers whose direct ray-tracing solution intercepts the array at an angle steeper than $-25^{\circ}$ from from horizontal; this can be seen in Ref.~\cite{Allison:2019rgg}, where there is a deficiency of triggers in A2 and A3 after the pulser is lowered below 1300m depth, despite the pulser being lowered to a total depth of 1700m. Therefore, for the calculation of $[A\Omega]_{\trm{eff}}$ used in the limit, we conservatively exclude neutrino simulated events with the same ray-tracing conditions.
This results in a ${\sim}10-30\%$ reduction in sensitivity, depending on energy. Excluding these steeply upgoing events is a conservative approach, as more exhaustive future studies might reveal that the cause of the trigger inefficiency to the calibration pulses does not have the same effect on neutrino events.

\begin{table}
\small
\centering
\begin{tabular}{p{35mm}| p{10mm} | p{10mm}}
\hline \hline{}
Systematic Uncertainty & + (\%) & - (\%) \\
\hline
Cross-Section (CTW) & 18 & 15 \\
Askaryan Emission & 13 & 11  \\
\hline\hline
Attenuation Length & 50 & 8 \\
Index of Refraction & 5 & 5 \\
Signal Chain & & 3 \\
Triggering Efficiency & & 3 \\
Total & 50 & 11\\
\hline
\end{tabular}
\caption{A summary of the systematic uncertainties in the neutrino sensitivity at a neutrino energy of $10^{18}$\,eV.}
\label{tab:systematic_sizes}
\end{table}

\section{Discussion and Outlook}
\label{sec:discuss}

In this paper, we present constraints on the flux of UHE neutrinos between $10^{16}$ and $10^{21}$\,eV from four years of data in A2 and A3. 
We have presented a description of the livetime and the instrument, and detailed the cuts used to eliminate backgrounds in two complementary, blind analyses.
The resultant limit from this search is the strongest limit set by ARA to date, and the strongest limit set by an in-ice radio neutrino detector above $10^{17}$\,eV.
The result utilizes more than quadruple the livetime of the previously published ARA analysis, and maintains reasonable efficiency to neutrinos while remaining general to signal shape and not requiring costly cuts on livetime in Austral summer or angular cuts in the direction of anthropogenic sources like South Pole Station. 
We are encouraged that the two analyses, which leveraged complimentary sets of reconstruction and analysis tools, have similar sensitivity and produced consistent expected limits within 15\% for all energy bins.

Post-unblinding, we were additionally able to further study our surface related backgrounds. As discussed previously, we observed zero events, consistent with our background estimates, including our estimated $10^{-3}$ events from above the surface. If we check the data taking runs that were excluded pre-unblinding because of the presence of large amounts of surface noise, we do observe a few events passing all cuts. We additionally are able to roughly estimate the probability of a single event of surface-origin being misreconstructed as coming from within the ice. To do so, we take the product of the fraction of runs in which we observe only one surface event and multiply by our estimated misreconstruction rate. We estimate the misreconstruction rate by taking the ratio of the number of events reconstructing inside the ice, relative to those reconstructing outside the ice, in the surface noisy runs. For example, in A3, we find that there may be approximately 0.2 such ``misreconstructing singlets." We note that this estimate is biased to larger values, because in order to measure the misreconstruction rate, we rely on the number of events reconstructing inside the ice in runs which demonstrate large amounts of surface noise, and were decided pre-unblinding to be unfit for analysis.  These two post-unblinding studies demonstrate the role of the surface-noisy cut in the present analysis, and represent an opportunity for growth during the development of future reconstruction techniques.

We underscore several important features of this newest result. First, it demonstrates ARA's capability to analyze its growing dataset. Compared to our previous result, which analyzed the first 10 months of data from stations A2 and A3 \cite{Allison:2015eky}, this analysis leverages data from four years of data-taking in each of the two stations. After removing intermittent periods of downtime we have about 1100 days (75\%) of livetime that was good for analysis for each station. This amounts to 2162 days of combined livetime.
This analysis is therefore the first ARA result to analyze $\mathcal{O}(10)$ station-years of data. This demonstrates the capability to analyze our growing dataset, which will be important as ARA looks to the future. 
There is roughly 4080 additional days of livetime awaiting analysis on archive, with the analysis pending ongoing calibration efforts.
With the full five-station ARA array collecting data since January 2018, the data set is expected to roughly double again
by 2022 (total of approximately 11k days of livetime). In Fig.~\ref{fig:limit} we additionally show the projected trigger-level single-event sensitivity that the five-station ARA5 array can achieve with data that will have been accumulated through 2022. As can be seen, ARA is poised to be the leading UHE neutrino detector above $10^{17}$~eV; the IceCube and Auger experiments will also accumulate additional livetime amounting to about 40\% and 25\% increases over their respective published limits.

Second, the analysis maintains reasonable efficiency (${\sim}35$\% at $10^{18}$\,eV, and reaching 50\% efficiency near a voltage signal-to-noise ratio (SNR) of 6) while remaining general and not relying on quantities that are
strongly model-dependent, such as a correlation with a signal template. This is advantageous because although the Askaryan signal has been observed in the laboratory {and in the atmosphere from cosmic-ray air showers~\cite{Belletoile:2015rea, Aab:2014esa, Scholten:2016gmj}, it has never before been observed in a dense media in nature.

In line with our previous two-station result \cite{Allison:2015eky}, this analysis did not require excluding data recorded during the Austral summer, nor did it require geometric rejection regions specifically in the direction of the South Pole. In the prior analyses of the Prototype station~\cite{Allison:2014kha,Allison:2015lnj}, 31\% of livetime was lost due to anthropogenic activities during the Austral summer, as well as 9\% due to the detector's solid angle coverage in directions near the South Pole.

We note three challenges overcome in these analyses that have resulted in improvements moving forward, especially as the ARA dataset continues to grow, the diversity of the array increases, and the field looks forward to a large-scale radio array in IceCube-Gen2.
The first challenge was managing the time-dependent nature of the ARA instruments. Some of the time dependent nature is owed to the different data taking configurations, as described in App.~\ref{app:livetime}, and often reflect improved understanding of the instrument and the ice. For example, an early trigger configuration led to triggering signals being off-center in the digitized waveform, and this was later corrected. The change to the readout length was made after Monte Carlo studies revealed that longer readout windows increased the probability with which a station records both the direct and refracted/reflected pulse that are possible because of the depth dependent index of refraction. Since learning from these processes, we have reached more stable operations configurations, and are working on additional streamlining. Some time dependence is owed to changing detector characteristics; for example, for some periods of time in ARA station 3, a digitization board exhibited a high amount of readout noise.
Such time-dependent detector features required adjustments to analysis algorithms and analysis thresholds. As a result of the analyses described herein, identification of such time periods has also been considerably streamlined. 

The second challenge was improvement in intra-collaboration communication between the ARA operations and analysis teams. In many cases, periods of livetime that were contaminated with calibration activity were recorded in operations reports, but were only later accounted for the analyses. We plan to work to streamline this pipeline for future ARA analyses. 

The third challenge was managing anthropogenic activity from the South Pole over several Austral summers. Despite most human activity being isolated nearly two miles away, the analysis requires aggressive cuts on downgoing signals, which eliminated 10-30\% of neutrino events. Improvements to reconstruction algorithms to more confidently reject downgoing events without requiring such substantial cuts on solid angle, or to more confidently reconstruct events with low hit-multiplicity, will improve the analysis efficiencies in the future.

\section{Acknowledgments}
The main authors of this manuscript were Brian Clark, Ming-Yuan Lu, and Jorge Torres, with Brian Clark and Ming-Yuan Lu leading the data analysis for this result. The ARA Collaboration designed, constructed and now operates the ARA detectors. Data processing and calibration, Monte Carlo simulations of the detector and of theoretical models and data analyses were performed by a large number of collaboration members, who also discussed and approved the scientific results presented here. We are grateful for contributions and discussions from Lucas Smith and Suren Gourapura.

We are thankful to the National Science Foundation (NSF) Office of Polar Programs and Physics Division for funding support through grants 1806923, 1404266, OPP-902483, OPP-1359535, and 1607555.
We further thank the Taiwan National Science Councils Vanguard Program NSC 92-2628-M-002-09 and the Belgian F.R.S.-FNRS Grant 4.4508.01.
We also thank the University of Wisconsin Alumni Research Foundation, the University of Maryland, and the Ohio State University for their support.
B.~A.~Clark thanks the NSF for support through the Graduate Research Fellowship Program Award DGE-1343012 and the Astronomy and Astrophysics Postdoctoral Fellowship under Award 1903885, as well as the Institute for Cyber-Enabled Research at Michigan State University.
A.~Connolly thanks the NSF for CAREER Award 1255557 and Award GRT00049285 and also the Ohio Supercomputer Center.
K.~Hoffman likewise thanks the NSF for their support through CAREER award 0847658.
S. A. Wissel thanks the NSF for support through CAREER Award 1752922 and the Bill and Linda Frost Fund at the California Polytechnic State University.
A.~Connolly, H.~Landsman, and D.~Besson thank the United States-Israel Binational Science Foundation for their support through Grant 2012077.
A.~Connolly, A.~Karle, and J.~Kelley thank the NSF for the support through BIGDATA Grant 1250720.
D.~Besson and A.~Novikov acknowledge support from National Research Nuclear University MEPhi (Moscow Engineering Physics Institute).
K.~Hughes thanks the NSF for support through the Graduate Research Fellowship Program Award DGE-1746045. 
A.~Vieregg thanks the Sloan Foundation and the Research Corporation for Science Advancement.
R.~Nichol thanks the Leverhulme Trust for their support.
K.D. de Vries is supported by European Research Council under the EU-ropean Unions Horizon 2020 research and innovation program (grant agreement No 805486).

Finally, we are thankful to the Raytheon Polar Services Corporation, Lockheed Martin, and the Antarctic Support Contractor for field support and enabling our work on the harshest continent.

\bibliographystyle{apsrev4-2}
\bibliography{references}

\appendix

\section{Limit Calculation}
\label{app:limit_calc}

We set a 90\% confidence level upper limit on the flux $E_{\trm{i}}F(E)_{\trm{i}}$ in the $i$-th energy bin of width $\textrm{dlog}_{10}E$ according to Equation~\ref{equ:limit_equation}:
\begin{equation}
    E_iF(E)_i = \frac{n_i}{\Lambda_i \,\,\, \ell n(10) \,\,\, \textrm{dlog}_{10}E}
    \label{equ:limit_equation}
\end{equation}
where $n_{\trm{i}}$ is the Feldman-Cousins upper limit for zero measured events on a background of  zero, accounting for the systematic uncertainties (added in quadrature) described in Sec.~\ref{sec:systematics}. This is done according to the prescription in Conrad~\textit{et.~al.}~\cite{Conrad:2002kn} with the improvements suggested by Hill~\textit{~et.~al.}~\cite{Hill:2003jk}.
Note that in the absence of uncertainties, $n_{\trm{i}}=2.44$, as commonly observed in the literature~\cite{Feldman:1997qc}.
Use of zero instead of the actual background estimate for the analysis setting the limit is conservative by $\sim2$\%, and does not substantially change the result.

We take $\textrm{dlog}_{10}E=1$, corresponding to decade wide bins in energy. $\Lambda_i$ is the exposure of the instrument summed over stations and configurations, taking into account analysis efficiencies as presented in Fig.~\ref{fig:efficiency}. $\Lambda_i$ for a given energy bin is defined explicitly as:
\begin{equation}
    \Lambda_i = \sum_{j_{\rm{stations}}=1}^{2} \sum_{k_{\rm{configs}}=1}^{5} \epsilon_{i,j,k} \;\; [A\Omega]_{\textrm{eff},i,j,k} \;\; T_{j,k}
\end{equation}
where for a specific energy ($i$), station ($j$), and configuration ($k$), $\epsilon_{\trm{i,j,k}}$ is the efficiency and $[A\Omega]_{\trm{eff,i,j,k}}$ is the effective area as shown in Fig.~\ref{fig:aeff}. $T_{\trm{j,k}}$ is the livetime of the instrument for a specific station and configuration as reported in Tab.~\ref{tab:configs}.

The effective areas $[A\Omega]_{\trm{eff}}$ of the instruments are computed from the effective volumes $[V\Omega]_{\trm{eff}}$ through the thin-target approximation:
\begin{equation}
    [A\Omega]_{\rm{eff}} = \frac{[V\Omega]_{\rm{eff}}}{\mathcal{L}_{\rm{int}}}
\end{equation}
where $\mathcal{L}_{\trm{int}}=m_N/(\rho ~ \sigma_{\nu-N})$ is the interaction length of a neutrino in the earth,

where $m_N = 1.67 \times 10^{-24} \textrm{g}$ is the mass of a nucleon, 
$\rho = 0.92 ~ \mathrm{g/cm^3}$ is the average density of ice, 
and $\sigma_{\nu-N}$ is the neutrino/anti-neutrino-nucleon cross-section in the units 
of $\mathrm{cm^2}$ as computed in Connolly~\textit{et.~al.}~\cite{Connolly:2011vc}.
The effective volumes are calculated by Monte Carlo methods. Using \texttt{AraSim} at specific energy, we simulate a number of neutrinos $N_{\trm{thrown}}$ in an interaction volume $V_{\trm{thrown}}$ with an isotropic distribution of arrival directions, and with an equal number of events between the three neutrino flavors and between neutrinos/anti-neutrinos. The sum of the weights of triggered events, $\sum w_{\trm{trig}}$,
determines the effective volume:

\begin{equation}
[V\Omega]_{\rm{eff}} = \frac{\sum w_{\rm{trig}}}{N_{\rm{thrown}}} \times V_{\rm{thrown}} \times 4\pi
\end{equation}
where the weighting accounts for the neutrino survival probability up to the interaction vertex.

\subsection{Limit with Alternative Flux Scaling}
\label{sec:limit_altscaling}
In Fig.~\ref{fig:limit_E2FE}, we show the same 90\% CL limit as we presented in Fig.~\ref{fig:limit}, but with an alternative scaling on the y-axis so that the flux is multiplied by an additional factor of energy in GeV. In this scaling, the y-axis represents \textit{energy} flux per $\textrm{cm}^2$$\textrm{s}^{-1}$$\textrm{sr}^{-1}$ as opposed to the \textit{particle} flux.
\begin{figure}[H]
\centering
\includegraphics[width=\columnwidth]{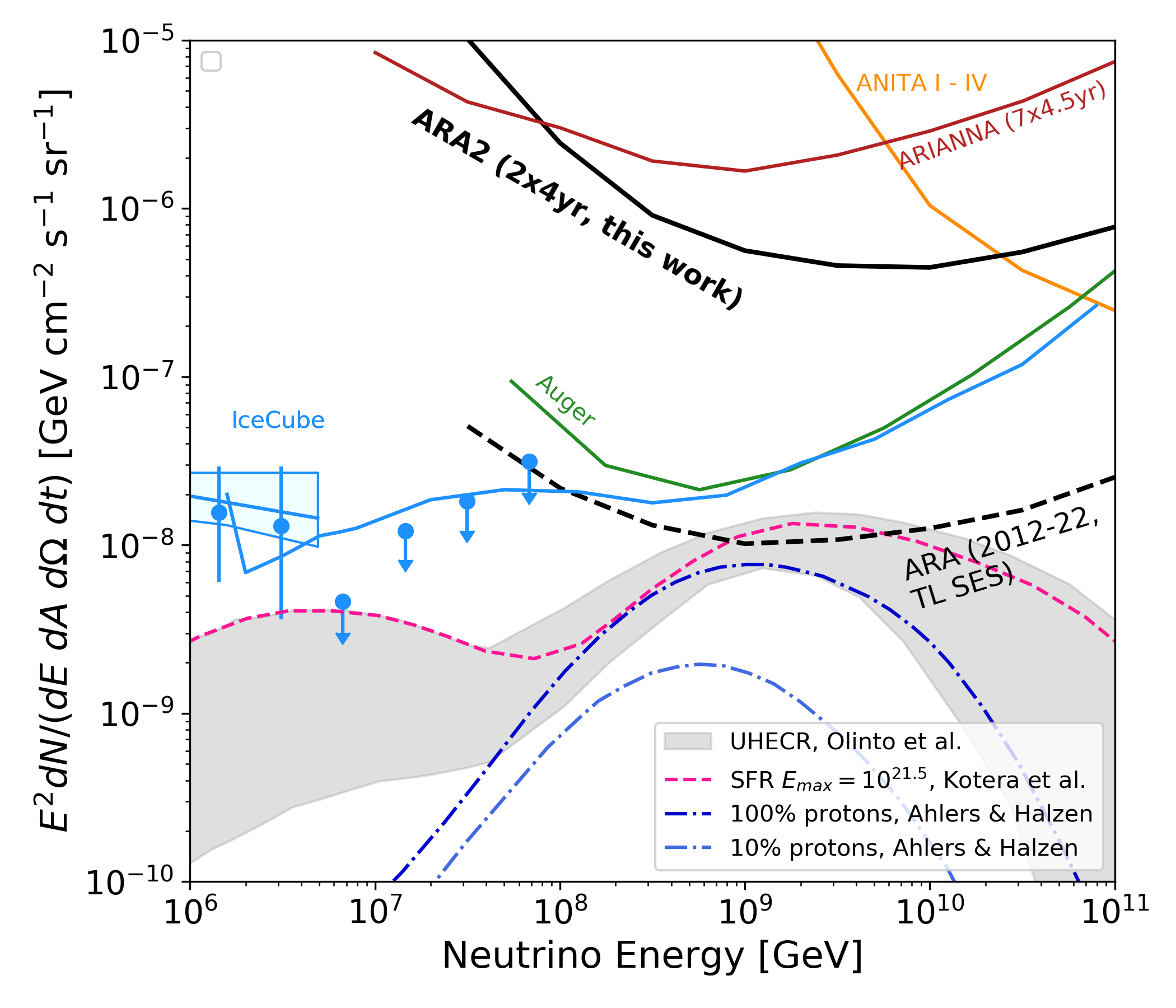}
\caption{Same as Fig.~\ref{fig:limit}, but with an alternative flux scaling.}
\label{fig:limit_E2FE}
\end{figure}

\subsection{Livetime Configurations}
\label{app:livetime}
In Tab.~\ref{tab:configs}, we provide a table
of the different station configurations, outlining the key parameters that differentiate them, along with the quantity of livetime for which they were active and the total number of events recorded (in millions). Note that A3 configuration 5 varies from A3 configuration 1 by the loss of quality digitizer data in string 4. The \textit{Trigger Window} is the amount of time during which 3/8 same-polarization antennas must have coincident single-channel triggers to trigger the readout of the instrument. The \textit{Readout Window} is the length of time for which the digitizers are read out. The \textit{Active Delays} column represents whether a set of trigger delays were applied to account for different cable lengths from different channels or not. 
\begin{table}
\begin{tabular}{|c|p{9mm}|p{11.75mm}|p{11.75mm}|p{9.5mm}|p{12mm}|p{12.5mm}|}
\hline
\multicolumn{1}{|l|}{Station} & Config & Readout Window (ns) & Trigger Window (ns) & Active Delays & Livetime (days) & Num Events (million) \\ \hline \hline
\multirow{5}{*}{2} & 1 & 400 & 110 & yes & 179 & 108.9 \\ \cline{2-7} 
                  & 2 & 400 & 110 & no  & 142 & 97.3\\ \cline{2-7} 
                  & 3 & 400 & 110 & yes & 94  & 54.5\\ \cline{2-7} 
                  & 4 & 520 & 170 & yes & 439 & 216.8\\ \cline{2-7} 
                  & 5 & 520 & 170 & no  & 287 & 129.7\\ \hline \hline
\multirow{5}{*}{3} & 1 & 400 & 110 & yes & 79 & 43.7\\ \cline{2-7} 
                  & 2 & 400 & 110 & no  & 147 & 114.2\\ \cline{2-7} 
                  & 3 & 520 & 170 & yes & 345 & 207.0\\ \cline{2-7} 
                  & 4 & 520 & 170 & no  & 260 & 171.8\\ \cline{2-7} 
                  & 5 & 400 & 110 & yes & 191 & 118.8\\ \hline
\end{tabular}
\caption{\label{tab:configs} Configuration definitions for A2 and A3, highlighting their various trigger parameters and livetimes.}
\end{table}

\begin{figure*}[hbt!]
        \centering
            \includegraphics[width=0.35\linewidth]{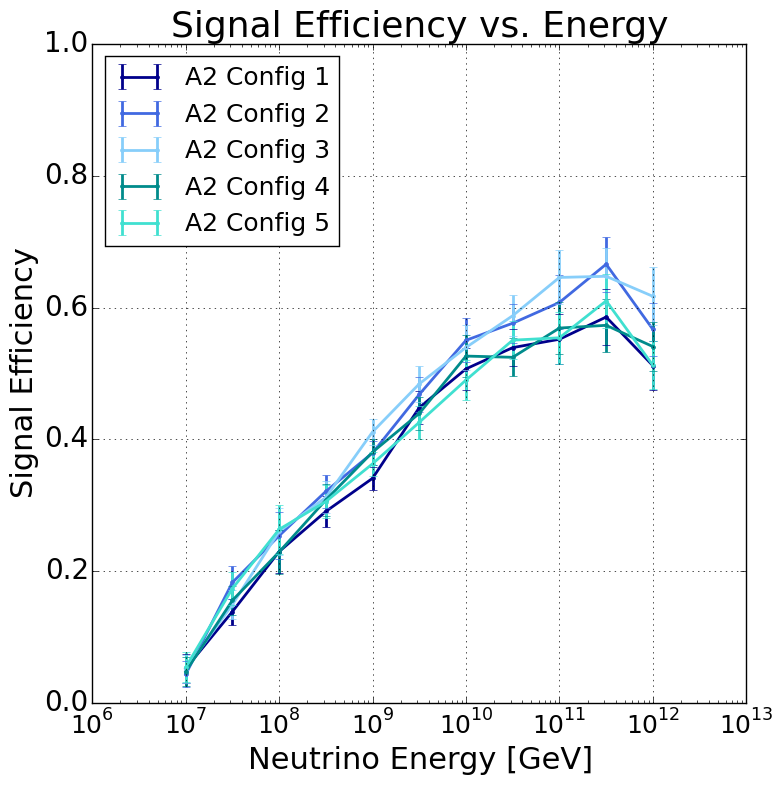}
            \label{fig:config_eff_a}
        \quad
            \centering 
            \includegraphics[width=0.35\linewidth]{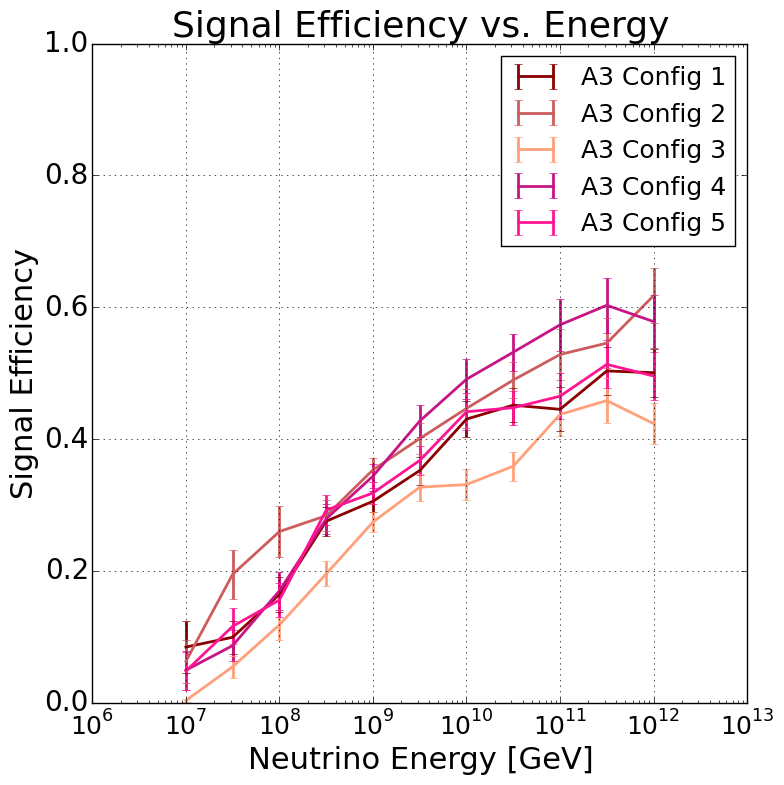}
            \label{fig:config_eff_b}
        \vskip\baselineskip
            \centering 
            \includegraphics[width=0.35\linewidth]{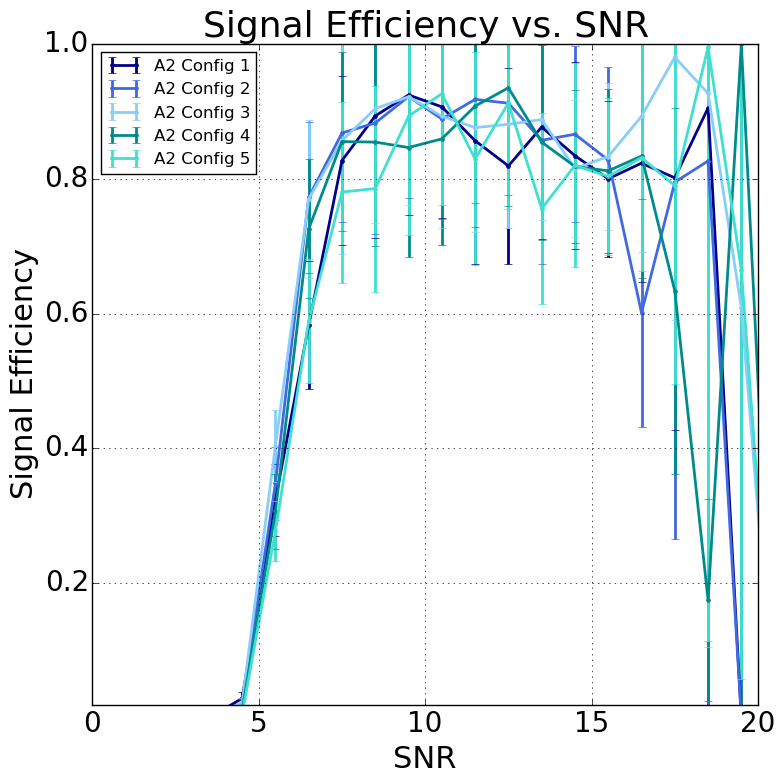}
            \label{fig:config_eff_c}
        \quad
            \centering 
            \includegraphics[width=0.35\linewidth]{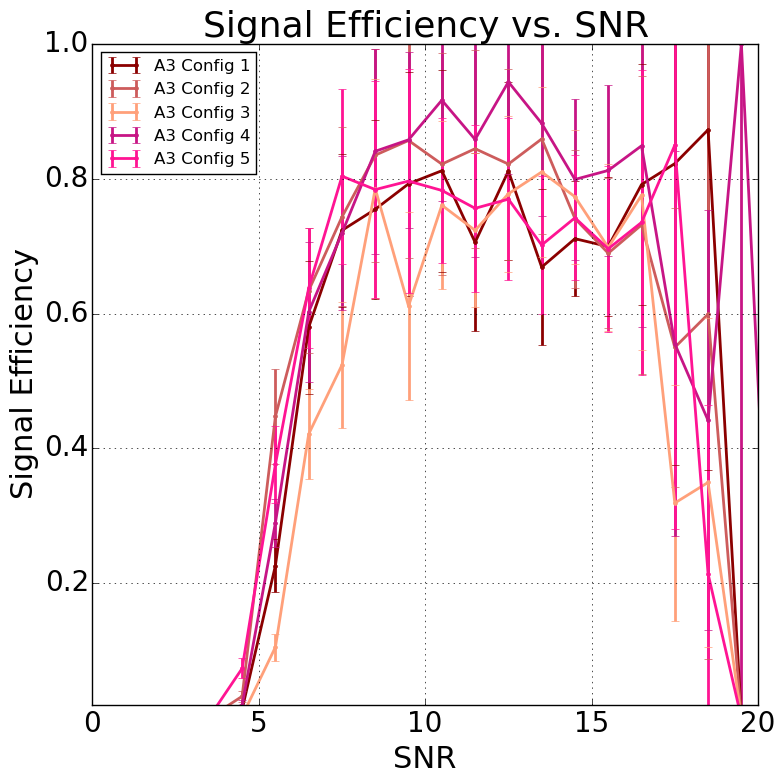}
            \label{fig:config_eff_d}
        \caption{Signal efficiency from Analysis A for each configuration in A2 and A3, respectively. Top row: signal efficiency as a function of neutrino energy for A2 (top left) and A3 (top right). Bottom row: signal efficiency as a function of signal-to-noise-ratio for A2 (bottom left) and A3 (bottom right) assuming the same energy weighting scheme as in Fig.~\ref{fig:efficiency}.
        } 
        \label{fig:config_eff}
    \end{figure*}

In Fig.~\ref{fig:config_eff}, we present the analysis efficiency from Analysis A for each configuration in both stations, as estimated using \texttt{AraSim} simulation. The analysis efficiency is driven by the analysis cuts, which depend on the background distribution observed during the livetime of each configuration. 

\end{document}

%% file: ara_revtex_institutes.tex

\newcommand{\atOSU}{\affiliation{Dept. of Physics, Center for Cosmology and AstroParticle Physics, The Ohio State University, Columbus, OH 43210}}
\newcommand{\atChiba}{\affiliation{Dept. of Physics, Chiba University, Chiba, Japan}}
\newcommand{\atUW}{\affiliation{Dept. of Physics, University of Wisconsin-Madison, Madison,  WI 53706}}
\newcommand{\atKU}{\affiliation{Dept. of Physics and Astronomy, University of Kansas, Lawrence, KS 66045}}
\newcommand{\atMoscow}{\affiliation{Moscow Engineering Physics Institute, Moscow, Russia}}
\newcommand{\atNTU}{\affiliation{Dept. of Physics, Grad. Inst. of Astrophys., Leung Center for Cosmology and Particle Astrophysics, National Taiwan University, Taipei, Taiwan}}
\newcommand{\atMSU}{\affiliation{Dept. of Physics and Astronomy, Michigan State University, East Lansing, Michigan 48824}}
\newcommand{\atUC}{\affiliation{Dept. of Physics, Enrico Fermi Institue, Kavli Institute for Cosmological Physics, University of Chicago, Chicago, IL 60637}}
\newcommand{\atUCL}{\affiliation{Dept. of Physics and Astronomy, University College London, London, United Kingdom}}
\newcommand{\atVUB}{\affiliation{Vrije Universiteit Brussel, Brussels, Belgium}}
\newcommand{\atUMD}{\affiliation{Dept. of Physics, University of Maryland, College Park, MD 20742}}
\newcommand{\atWhittier}{\affiliation{Dept. Physics and Astronomy, Whittier College, Whittier, CA 90602}}
\newcommand{\atUNL}{\affiliation{Dept. of Physics and Astronomy, University of Nebraska, Lincoln, Nebraska 68588}}
\newcommand{\atWeizman}{\affiliation{Weizmann Institute of Science, Rehovot, Israel}}
\newcommand{\atUD}{\affiliation{Dept. of Physics, University of Delaware, Newark, DE 19716}}
\newcommand{\atDenison}{\affiliation{Dept. of Physics and Astronomy, Denison University, Granville, Ohio 43023}}
\newcommand{\atUH}{\affiliation{Dept. of Physics and Astronomy, University of Hawaii, Manoa, HI 96822}}
\newcommand{\atCalPoly}{\affiliation{Physics Dept., California Polytechnic State University, San Luis Obispo, CA 93407}}
\newcommand{\atPSU}{\affiliation{Dept. of Physics, Dept. of Astronomy and Astrophysics, Pennsylvania State University, State College, PA 16802}}

%% file: ara_revtex_authors.tex

 \author{P.~Allison}\atOSU
 \author{S.~Archambault}\atChiba
 \author{J.J.~Beatty}\atOSU
 \author{M.~Beheler-Amass}\atUW
 \author{D.Z.~Besson}\atKU\atMoscow
 \author{M.~Beydler}\atUW
 \author{C.C.~Chen}\atNTU
 \author{C.H.~Chen}\atNTU
 \author{P.~Chen}\atNTU
 \author{B.A.~Clark}\email[B. A. Clark: ]{baclark@msu.edu}\atOSU\atMSU
 \author{W.~Clay}\atUC
 \author{A.~Connolly}\atOSU
 \author{L.~Cremonesi}\atUCL
 \author{J.~Davies}\atUCL
 \author{S.~de~Kockere}\atVUB
 \author{K.D.~de~Vries}\atVUB
 \author{C.~Deaconu}\atUC
 \author{M.A.~DuVernois}\atUW
 \author{E.~Friedman}\atUMD
 \author{R.~Gaior}\atChiba
 \author{J.~Hanson}\atWhittier
 \author{K.~Hanson}\atUW
 \author{K.D.~Hoffman}\atUMD
 \author{B.~Hokanson-Fasig}\atUW
 \author{E.~Hong}\atOSU
 \author{S.Y.~Hsu}\atNTU
 \author{L.~Hu}\atNTU
 \author{J.J.~Huang}\atNTU
 \author{M.-H.~Huang}\atNTU
 \author{K.~Hughes}\atUC
 \author{A.~Ishihara}\atChiba
 \author{A.~Karle}\atUW
 \author{J.L.~Kelley}\atUW
 \author{R.~Khandelwal}\atUW
 \author{K.-C.~Kim}\atUMD
 \author{M.-C.~Kim}\atChiba
 \author{I.~Kravchenko}\atUNL
 \author{K.~Kurusu}\atChiba
 \author{H.~Landsman}\atWeizman
 \author{U.A.~Latif}\atKU
 \author{A.~Laundrie}\atUW
 \author{C.-J.~Li}\atNTU
 \author{T.-C.~Liu}\atNTU
 \author{M.-Y.~Lu}\email[M.-Y. Lu: ]{mlu27@wisc.edu}\atUW 
 \author{B.~Madison}\atKU
 \author{K.~Mase}\atChiba
 \author{T.~Meures}\atUW
 \author{J.~Nam}\atChiba
 \author{R.J.~Nichol}\atUCL
 \author{G.~Nir}\atWeizman
 \author{A.~Novikov}\atKU\atMoscow
 \author{A.~Nozdrina}\atKU
 \author{E.~Oberla}\atUC
 \author{A.~O'Murchadha}\atUW
 \author{J.~Osborn}\atUNL
 \author{Y.~Pan}\atUD
 \author{C.~Pfendner}\atDenison
 \author{J.~Roth}\atUD
 \author{P.~Sandstrom}\atUW
 \author{D.~Seckel}\atUD
 \author{Y.-S.~Shiao}\atNTU
 \author{A.~Shultz}\atKU
 \author{D.~Smith}\atUC
 \author{J.~Torres}\email[J. Torres: ]{torresespinosa.1@osu.edu}\atOSU
 \author{J.~Touart}\atUMD
 \author{N.~van~Eijndhoven}\atVUB
 \author{G.S.~Varner}\atUH
 \author{A.G.~Vieregg}\atUC
 \author{M.-Z.~Wang}\atNTU
 \author{S.-H.~Wang}\atNTU
 \author{S.A.~Wissel}\atCalPoly\atPSU
 \author{S.~Yoshida}\atChiba
 \author{R.~Young}\atKU
\collaboration{ARA Collaboration}\noaffiliation